\documentclass{jfm}
\usepackage{graphicx}
\usepackage{epstopdf, epsfig}
\usepackage{amsmath,bm}
\usepackage{amssymb}
\usepackage{graphicx}
\usepackage{subcaption}
\usepackage[dvipsnames]{xcolor}
\usepackage[normalem]{ulem}
\usepackage{float}

\newcommand{\obu}{\bm{\overline{u}}}
\newcommand{\hbu}{\bm{\hat{u}}}
\newcommand{\hbf}{\bm{\hat{f}}}

\newcommand{\hbd}{\bm{\hat{d}}}
\newcommand{\hbq}{\bm{\hat{q}}}
\newcommand{\hf}{\hat{f}}
\newcommand{\hu}{\hat{u}}
\newcommand{\tq}{\bm{q}}
\newcommand{\td}{\bm{d}}
\newcommand{\htq}{\hat{\bm{q}}}
\newcommand{\htd}{\hat{\bm{d}}}

\newcommand{\wzm}{\omega_{o,m}}
\newcommand{\ezm}{\epsilon_{o,m}}
\newcommand{\Ri}{R(i\wz)^{-1}}
\newcommand{\e}{\epsilon}
\newcommand{\ez}{\epsilon_o}
\newcommand{\wz}{\omega_o}
\newcommand{\de}{\delta}
\newcommand{\w}{\omega}
\newcommand{\forcutwo}{\tilde{\bm{u}}_2}
\newcommand{\forcuthree}{\tilde{\bm{u}}_3}
\newcommand{\bu}{\bm{u}}
\newcommand{\bff}{\bm{f}}
\newcommand{\bv}{\bm{v}}

\newcommand{\bl}{\bm{l}}

\newcommand{\hL}{L}
\newcommand{\ba}{\bm{b}}
\newcommand{\mL}{\mathcal{L}}
\newcommand{\da}{\dagger}
\newcommand{\md}{\mathrm{d}}

\newcommand{\bs}{\bm{s}}
\newcommand{\Heq}{\left | a_e \right |}
\newcommand{\Peq}{\rho_e}
\newcommand{\ord}{O}
\newcommand{\hba}{\bm{\hat{a}}}
\newcommand{\bq}{\bm{q}}
\newcommand{\pa}{\partial}
\newcommand{\ampa}{\left | a \right |}

\shorttitle{Weak nonlinearity for strong nonnormality}
\shortauthor{Y.-M. Ducimeti\`ere, E. Boujo and F. Gallaire}

\title{Weak nonlinearity for strong nonnormality}

\author{Yves-Marie Ducimeti\`ere\aff{1}
  \corresp{\email{yves-marie.ducimetiere@epfl.ch}},
  Edouard Boujo\aff{1}
 \and Fran\c cois Gallaire\aff{1}}

\affiliation{\aff{1}Laboratory of Fluid Mechanics and Instabilities, EPFL, CH1015 Lausanne, Switzerland}

\begin{document}

\maketitle

\begin{abstract}
We propose a theoretical approach to derive amplitude equations governing the weakly nonlinear evolution of nonnormal dynamical systems, when they experience transient growth or respond to harmonic forcing. This approach reconciles the nonmodal nature of these growth mechanisms and the need for a centre manifold to project the leading-order dynamics. 
Under the hypothesis of strong nonnormality, we take advantage of the fact that small operator perturbations suffice to make the inverse resolvent and the inverse propagator singular, which we encompass in a multiple-scale asymptotic expansion.  
The methodology is outlined for a generic nonlinear dynamical system, and four application cases highlight common nonnormal mechanisms in hydrodynamics: the streamwise convective nonnormal amplification in the flow past a backward-facing step, and the Orr and lift-up mechanisms in the plane Poiseuille flow. 
\end{abstract}

\begin{keywords}
keywords
\end{keywords}

\section{Introduction}

Nonlinear dynamical systems can have one or several equilibrium solutions, which form one of the building blocks of the phase space \cite{Strogatz15}. The linear stability of an equilibrium  can be deduced from the eigenvalues of the linearised operator: linear modal analysis thus helps to distinguish between linearly \textit{unstable}, \textit{neutral}  (\textit{marginally stable}) and \textit{strictly stable} equilibria (when the largest growth rate is positive, null and negative, respectively), and to detect bifurcations.
It sometimes remains too simplistic, however, and  has therefore been generalised over the last decades  to account for nonlinear and nonmodal effects, although these two types of correction have generally been opposed \cite{Trefethen93}, culminating into a paper entitled ``\textit{Nonlinear normality versus non-normal linearity}'' \cite{Waleffe95}. 
The objective of the present study is precisely to contribute to reconcile nonlinearity and nonnormality, and to rigorously derive weakly nonlinear amplitude equations ruling nonnormal systems.

\subsection{Weak nonlinearity}

While the most unstable eigenmode eventually dominates the linear, small-amplitude dynamics, its shape and frequency may differ significantly from those of the nonlinear state when moving away from the bifurcation point. 
Fundamentally, the saturation amplitude can only be  determined trough nonlinear considerations, and one must resort to a weakly or fully nonlinear analysis. 
Following the insight of Lev Landau, who introduced amplitude equations in analogy to phase transitions (\cite{Landau87}, §26), weakly nonlinear analyses using a multiple-scale approach leading to an equation for the bifurcated mode amplitude $A$ were performed in some pioneering works in the context of thermal convection  \cite{Gor57,Malkus58},  parallel shear flows \cite{Stuart58, Stuart60, Watson60} and non-parallel \cite{Sipp07} shear flows.
In theses studies, a so-called Stuart-Landau equation of the form $\mathrm{d}_T A = \lambda A - \kappa A \left | A \right |^2$ is obtained as a condition for non-resonance. When the real part of the nonlinear coefficient is strictly positive, $\Re(\kappa)>0$, the cubic term $A\left | A \right |^2$ is sufficient to capture the saturation amplitude, and the Stuart-Landau equation is an accurate model for supercritical bifurcations; otherwise it can be extended to describe subcritical bifurcations. Amplitude equations, which can also depend on space, are widely used to describe the spatiotemporal pattern formation in physical systems near threshold. 
Beyond hydrodynamics, this occurs in plasma physics, solidification fronts, nonlinear optics, laser physics, oscillatory chemical reactions, buckling of elastic rods, and many others fields of study (see \cite{Cross93} for a comprehensive review). More generally, while the form of the amplitude equation can often be deduced from symmetry considerations \cite{Fauve98,Crawford91}, its coefficients ($\lambda$ and $\kappa$ in the case of the Stuart-Landau equation) are evaluated with  scalar products of fields computed at the bifurcation point.

Other approaches exist to deduce the normal form, i.e. the amplitude equation which distillates the quintessence  of the nonlinear behaviour in the vicinity of a bifurcation point \cite{Manneville04, Guckenheimer83,Haragus11}. Common to all these approaches is the concept of centre manifold, along which the dynamics are slow, while, under a spectral gap assumption, an adiabatic elimination ensures the slaving of damped modes.

Regardless of whether the system is forced or freely evolving, an amplitude equation can only be constructed  close to a bifurcation point. Indeed, only linearised systems with a neutral or weakly damped eigenmode may experience resonance, whose avoidance condition results in the amplitude equation. 
Not all systems possess such eigenmodes, and systems with a significantly damped spectrum are often encountered. Nevertheless, such systems are still of great importance in practice, owing to so-called nonnormal amplifying mechanisms.


\subsection{Strong nonnormality}
\begin{figure}
\centering
\begin{subfigure}{0.49\linewidth}
\includegraphics[trim={11cm 20cm 1.75cm 1cm},clip,width=1\linewidth]{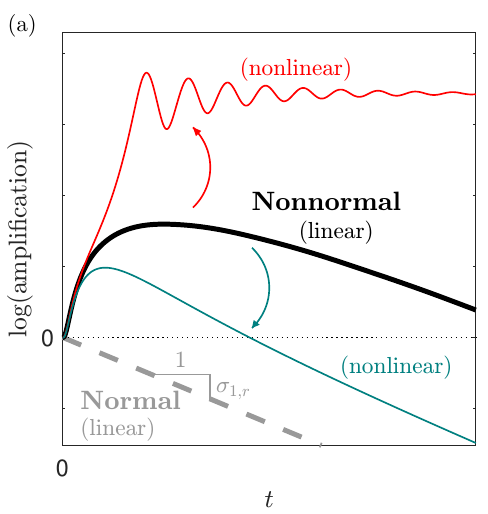}
\end{subfigure}
\hfill
\begin{subfigure}{0.49\linewidth}
\includegraphics[trim={11cm 5.5cm 1.75cm 10cm},clip,width=1\linewidth]{figures/sketch.pdf}
\end{subfigure}
\caption{Cartoon representation of nonlinearity and nonnormality, illustrated in the time domain (a) and frequency domain (b), for a linearly stable system; the least stable eigenvalue $\sigma_1$ of the eigenspectrum in (c) has indeed a negative growth rate. 
\textbf{(a)}~In the linear regime, the amplitude of the perturbations eventually decays like $\exp(\sigma_{1,r}t)$. Nonnormal systems can experience a very large transient growth. Nonlinearity may be stabilising or destabilising. 
\textbf{(b)}~Normal systems subject to external forcing respond preferentially at frequency $\sigma_{1,i}$. Nonnormal systems can respond at different frequencies, with an amplification much larger than predicted by $\sigma_{1,r}$. Nonlinearity may be stabilising or destabilising.}\label{fig:sketch}
\end{figure}

Upon the choice of a scalar product, a linear operator is nonnormal if it does not commute with its adjoint. Consequently, its eigenmodes do not form an orthogonal set, and the response to an initial condition and to time-harmonic forcing may be highly non-trivial (see \cite{Tref05} for an exhaustive presentation). This response generally results from an intricate cooperation between a large amount of eigenmodes. 
The leading (least stable or most unstable) eigenvalue solely provides the asymptotic (long-time) linear behaviour of the energy of the unforced system. At finite time, restriction to the leading eigenmode is generally irrelevant. In particular, a negative growth rate for all eigenvalues is not a guarantee for the energy to decay monotonously for all initial conditions: some small-amplitude perturbations may experience a large transient amplification (figure~\ref{fig:sketch}a).
The same is true for systems subject to harmonic forcing: they may exhibit strong amplification, much larger than the inverse of the smallest damping rate, and at forcing frequencies unpredictable at the sight of the spectrum (figure~\ref{fig:sketch}b).

Nonnormal operators are traditionally
encountered in laser physics (see \cite{Tref05} §60), which H. J. Landau  described by developing  the concept of \textit{pseudospectrum}, as a pertinent alternative to modal analysis \cite{Landau76,Landau77}. The physical implication of nonnormality in the unstable laser cavity is profound, as it results in a substantial increase in the linewidth of the laser beam signal compared to a perfect resonator \cite{Petermann79}. In astrophysics,  pseudospectrum analysis was used very recently to study the stability of black holes \cite{Jaramillo21}. 
In network science, a recent study \cite{Asllani18} reports a systematic analysis of a large set of (directed) empirical networks from a variety of disciplines including ``\textit{biology, sociology, communication, transport, and many more}", and gives evidence that all of them present a strong nonnormality. 
Nonnormality may shrink the basin of attraction of a linearly strictly stable equilibrium, as strong amplification may trigger nonlinearities  (figure~\ref{fig:sketch}), and radically change the behaviour of dynamical systems. 
This is illustrated in \cite{Asllani18b}, where the nonnormality of the London Tube network results in an hypothetical outbreak of  measles epidemic, although the linear stability theory predicts an asymptotic decay of the number of contagions. 
In hydrodynamics, nonnormality is frequent and inherited from the linearisation of the advective term $(\bm{U} \cdot \nabla) \bm{U}$.
This term gives a preferential direction to the fluid flow, which breaks the normality of the linear operator. 
In the context of parallel flows, nonnormality is found for instance in the canonical plane Couette and Poiseuille flows \cite{Gustavsson91, Butler92, Reddy93, Trefethen93, Farrell93, SH01}, pipe flow \cite{Schmid94}
and parallel boundary layers \cite{Butler92, Corbett00}. 
Nonnormality is also found in nonparallel flows \cite{Cossu97}, for instance spatially developing boundary layers \cite{Ehrenstein05, Akervik08, Ehrenstein08, Alizard09, Monkrousos10}, jets \cite{Garnaud13, Garnaud13B} and
the flow past a backward-facing step \cite{Blackburn08, Boujo15}. Exhaustive reviews of nonnormality in hydrodynamics can be found in \cite{Chomaz05, Schmid07}. The crucial role played by nonnormality in the transition to turbulence has become clear over the years \cite{Trefethen93, Baggett96, Schmid07}. As mentioned in the context of nonnormal networks, if the flow is nonnormal, low-energy perturbations such as free-stream turbulence or wall roughness can be amplified strongly enough to lead to a  regime where nonlinearities come into play, which may lead to turbulence through a sub-critical bifurcation. The toy system presented in \cite{Trefethen93} is an excellent illustration of this so-called ``bypass'' scenario.

\subsection{Amplitude equations without eigenvalues}
Through nonnormality, systems with strongly damped spectra can bear strong amplification of specific structures at relatively selective frequencies and/or temporal horizons. To the best of the authors' knowledge, it is currently impossible to construct an amplitude equation for such systems, again because no neutral bifurcation point exists. Furthermore, a systems may well have a weakly damped mode and still exhibit nonnormality, which would jeopardise a classical, single-mode amplitude equation.

Notwithstanding the relevance and usefulness of fully nonlinear solutions \cite{Hof04, Schneider10}, as well as the existence of a fully nonlinear nonnormal stability theory able to compute nonlinear optimal initial conditions via Lagrangian optimisation \cite{Cherubini10, Cherubini11, Pringle10, KerswellAnnuRev2018}, we believe that establishing a rigorous reduced-order model for weak nonlinearities is relevant. In specific regimes, such a model would quantify the respective contribution of each dominant nonlinear interaction, thus bringing insight on the saturation mechanisms of harmonic and transient amplification. It would also predict efficiently if such a saturation actually exists or if, on the contrary, nonlinearities tend to yield even stronger amplification than in the linear regime, thus leading to sub-critical or non-monotonous behaviors; the sketch in figure~\ref{fig:sketch} illustrates possible scenarios for the effects of nonlinearities subsequent to strong nonnormal amplification.
Finally, amplitude equations are useful for flow control and optimisation, as shown for instance in \cite{Sipp12} in the more classical context of a marginally stable flow, displaying little nonnormality, a well isolated eigenvalue and a sufficiently large spectral gap. 

\begin{figure*}
\centering
\includegraphics[trim={0.5cm 26.4cm 10.5cm 0.8cm},clip,width=1\linewidth]{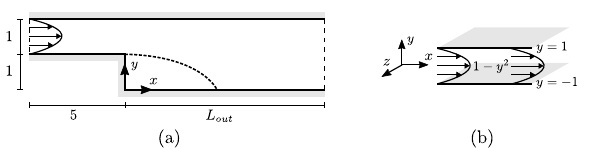}
\caption{Sketch of the flow configurations. \textbf{(a)}~\label{fig:sbfs}Two-dimensional flow over a backward-facing step, with fully developed parabolic profile of unit maximum centerline velocity at the inlet.
\textbf{(b)}~\label{fig:spois} Three-dimensional  plane Poiseuille flow, confined between two solid walls at $y=\pm1$, and invariant in the $x$ (streamwise) and $z$ (spanwise) directions}
\end{figure*}

The present work proposes to reconcile amplitude equations and nonnormality. Specifically, a method is advanced to derive amplitude equations in the context of (i)~harmonic forcing and (ii)~transient growth. 
In case (i), we vary the amplitude of a given harmonic forcing at a prescribed frequency and predict the gain (energy growth) of the asymptotic response (\S\ref{section:harmonic}). 
In case (ii), we vary the amplitude of a given initial condition and predict the gain of the response at a selected time $t=t_o$ (\S\ref{section:tg}).
In both cases, we perform an \textit{a priori} weakly nonlinear prolongation of the gain, at very low numerical cost. The applied harmonic forcing and initial condition are allowed to be arbitrarily different from any eigenmode. The method does not rely on the presence of an eigenvalue close to the neutral axis; instead, it applies to \textit{any} sufficiently nonnormal operator. If such an eigenvalue was nevertheless present on the neutral axis, we recover a classical, modal amplitude equation. The method is illustrated with two  flows, the nonparallel flow past a backward-facing step (sketched in figure~\ref{fig:sbfs}a) and the parallel plane Poiseuille flow (figure~\ref{fig:sbfs}b).  These two nonnormal flows  exhibit large gains, both in the context of harmonic forcing (\S\ref{sec:harmonicBFS}-\ref{sec:harmonicPoiseuille}) and transient growth (\S\ref{sec:tgBFS}-\ref{sec:tgPoiseuille}).

In both contexts, a generic nonlinear dynamical system is considered,
\begin{equation} 
\partial_t \bm{U}  =  N(\bm{U})  + \bm{F}, 
 \quad \quad
 \bm{U}(0) = \bm{U}_0, \label{eq:ieq}
\end{equation} 
where $N(*)$ is a nonlinear operator and $\bm{F}$ is a forcing term. 
An appropriate and common place to begin the analysis of  (\ref{eq:ieq}) is to linearise it around an unforced equilibrium. 
The latter is denoted $\bm{U}_e $ and satisfies $N(\bm{U}_e)=\bm{0}$. 
Around this equilibrium are considered small-amplitude perturbations in velocity $\e \bm{u}$, forcing $\e \bm{f}$, and initial condition $\e\bm{u}_0$, where $\e \ll 1$. An asymptotic expansion of (\ref{eq:ieq}) in terms of $\e$ can thus be performed, transforming the nonlinear equation into a succession of linear ones. The fields $\bm{u}$, $\bm{f}$ and $\bm{u}_0$ are recovered at order $\e$ and linked trough the linear relation
\begin{equation}
\partial_t  \bm{u} = L\bm{u} + \bm{f}, 
\quad  
\bm{u}(0) = \bm{u}_0,
\label{eq:ieqlin}
\end{equation}
where $L$ results from the linearisation of $N$ around $\bm{U}_e$. For fluid flows governed by the incompressible Navier-Stokes equations, $L \bm{u}  = - (\bm{U}_e \cdot \nabla) \bm{u} -(\bm{u} \cdot \nabla)\bm{U}_e + Re^{-1}\Delta \bm{u} - \nabla p(\bm{u})$, where the pressure field $p$ is such that the velocity field $\bm{u}$ is divergence-free. Both fields are linked trough a linear Poisson equation. In practice,  pressure is included in the state variable, resulting in a singular mass matrix; it is omitted here, for the sake of clarity.

\section{Response to Harmonic Forcing}
\label{section:harmonic}

We first derive an amplitude equation for the weakly nonlinear  amplification of time-harmonic forcing 
$\bm{f}(\bm{x},t) = \bm{\hat{f}}(\bm{x}) e^{i\wz t} +c.c$ in a linearly strictly stable system. In the long-time regime, only the same-frequency harmonic response $\bm{u}(\bm{x},t) = \bm{\hat{u}}(\bm{x}) e^{i\wz t} + c.c$ persists. Injecting the  expressions of $\bm{f}$ and $\bm{u}$ in (\ref{eq:ieqlin}) leads to $\bm{\hat{u}} = (i \wz I  - L)^{-1} \bm{\hat{f}} \doteq  R(i\wz) \bm{\hat{f}}$, where $R(z) = (z I  - L )^{-1}$ is the \textit{resolvent} operator. 
In the current context, it maps a harmonic forcing structure onto its asymptotic linear response at the same frequency. 
A measure of the maximum gain is 
\begin{equation}
\centering
G(i \wz)  =  \max_{\bm{\hat{f}}} \frac{\left \|  \bm{\hat{u}} \right \|}{\left \|  \bm{\hat{f}} \right \|}  
=  \left \| R(i \wz) \right \| 
\doteq \frac{1}{\ez}. 
\label{eq:hgain}
\end{equation}
In the following, we choose the $L^2$ norm (or ``energy'' norm) induced by the Hermitian inner product 
$\left \langle \hbu_a, \hbu_b \right \rangle 
= \int_{\Omega}^{} \hbu_a^H\hbu_b \mathrm{d}\Omega$ (the superscript $H$  denotes the Hermitian transpose). The operator $R(i \wz)^{\da}$ denotes the adjoint of $R(i \wz)$ under this scalar product, such that $\left \langle R(i \wz) \hbu_a,\hbu_b \right \rangle = \left \langle \hbu_a,R(i \wz)^{\da} \hbu_b \right \rangle$, for any $\hbu_a, \hbu_b$.
Among all frequencies $\wz$, the one leading to the \textit{maximum} amplification is noted $\wzm$ and associated with an optimal gain $G(i\wzm)=1/\ezm$. 
The \textit{singular value decomposition} of $R(i\wz)$ provides $G(i\wz) = \ez^{-1}$ as the largest singular value, and the associated pair of right singular vector $\hbf_o$ and left singular vector $\hbu_o$. The former represents the optimal forcing, whereas the latter characterises the long-time harmonic response reached, after the transients fade away: 
\begin{equation} 
\Ri\hbu_o = \ez \hbf_o,  
\quad 
\left[R(i\wz)^\da\right]^{-1}\hbf_o = \ez \hbu_o, \label{eq:1}
\end{equation}
where $||\hbf_o|| = ||\hbu_o || = 1$. 
Smaller singular values of $R(i\wz)$ constitute \textit{sub-optimal} gains, and the associated right singular vectors are sub-optimal forcing structures. 
Note that one can express $\langle \hbu , \hbu \rangle = \langle R\hbf , R\hbf \rangle $ as $ \langle R^{\da} R \hbf , \hbf \rangle$, such that the singular values of $R(i\wz)$ are also the square root of the eigenvalues of the \textit{symmetric} operator $R(i\wz)^{\da} R(i\wz)$. An important implication is that the singular vectors form an orthogonal set for the scalar product $\langle*,*\rangle$. 
The practical computation of $\ez$, $\hbf_o$ and $\hbu_o$ is detailed for the Navier-Stokes equations in \cite{Garnaud13B}, for instance. Note that if the operator $L$ possesses a neutral eigenvalue, $\wzm$, $\hbf_o$ and $\hbu_o$ respectively reduce to the frequency, the adjoint and the direct mode associated to this eigenvalue.

Since $L$ is strongly nonnormal, as assumed in the rest of the present study, none of $\ez$, $\hbu_o$ and $\hbf_o$ are immediately determined from its spectral (modal) properties. 
Strong nonnormality implies $\ez \ll 1$, such that the inverse resolvent $\Ri$ appearing in (\ref{eq:1}) is \textit{almost} singular. Perturbing it as  
\begin{equation}
\begin{split}
\Phi \doteq \Ri -\ez P, 
\quad \text{where} \quad  
P = \hbf_o \left \langle \hbu_o , \ast   \right \rangle, 
 \label{eq:pinvr}
\end{split}
\end{equation}
%
leads to $\Phi\hbu_o = \bm{0}$, such that $\Phi$ is \textit{exactly} singular. The norm of the perturbation operator is small since $||P||=1$. The field $\hbu_o$ constitutes the only non-trivial part of the kernel of $\Phi$, and its associated adjoint mode is $\hbf_o$. Indeed, using that $P^\da = \hbu_o\langle \hbf_o,* \rangle$, we have
\begin{align*}
\begin{split}
\Phi^\da \hbf_o &= \left[ \Ri \right]^\da \hbf_o -  \ez \hbu_o \left \langle \hbf_o, \hbf_o \right \rangle \\
&= \left[ R(i\wz)^\da \right] ^{-1} \hbf_o -  \ez \hbu_o   = \bm{0},
\end{split}
\end{align*}
where we used the fact that the inverse of the adjoint is the adjoint of the inverse.
We note that $\Phi$ can be rewritten as $\Phi = (i\wz I-\hL_n)$ where $\hL_n \doteq L + \ez P$, such that (\ref{eq:pinvr}) seems to imply that the state operator $L$ has been perturbed. In this process, the operator $\hL_n$ has acquired an eigenvalue equal to $i\wz$, and therefore has become  neutral. 
However, it has also lost its reality and therefore does not, in general, possess an eigenvalue equal to $-i\wz$. 
By construction, $\ez$ is the \textit{smallest} possible amplitude of the right-hand side of (\ref{eq:1}) for a given $i \wz$, such that $\ez P$ is the \textit{smallest}  perturbation of $L$ necessary to relocate an eigenvalue of $L$ on $i\wz$. This fact can be formalised with the pseudospectrum theory outlined in \cite{Tref05}. In the complex plane, $z \in \mathbb{C}$ belongs to the $\epsilon$-\textit{pseudospectrum} $\Lambda_{\epsilon}(L)$ if and only if $\left \| R(z) \right \| \geq 1/\epsilon$. 
If $E$ is an operator with $||E||=1$, eigenvalues of $L-\epsilon E$ can lie anywhere inside $\Lambda_{\epsilon}(L)$. 
Eigenvalues of $L$ and singularities of $\left \| R(z) \right \|$ thus collide with the $\e$-pseudospectrum in the limit $\epsilon \rightarrow 0$. 
As $\epsilon$ increases, the $\e$-pseudospectrum may touch the imaginary axis, such that any $z = i \wz $ can be an eigenvalue of $L-\epsilon E$ if the amplitude of the perturbation is greater than or equal to $\e = \left \| R(i \wz) \right \|^{-1}$. 
We recognise $\e$ as the inverse gain $\ez$ defined in (\ref{eq:hgain}), and thus $E$ as $P$. In particular, if $\wz = \wzm$, the associated $\ezm$ is referred to as the \textit{stability radius} of $L$ since the $\ezm$-pseudospectrum is the first to touch the imaginary axis.

\begin{figure}
\centering
\includegraphics[trim={3.3cm 10cm 4cm 10cm},clip,width=0.65\linewidth]{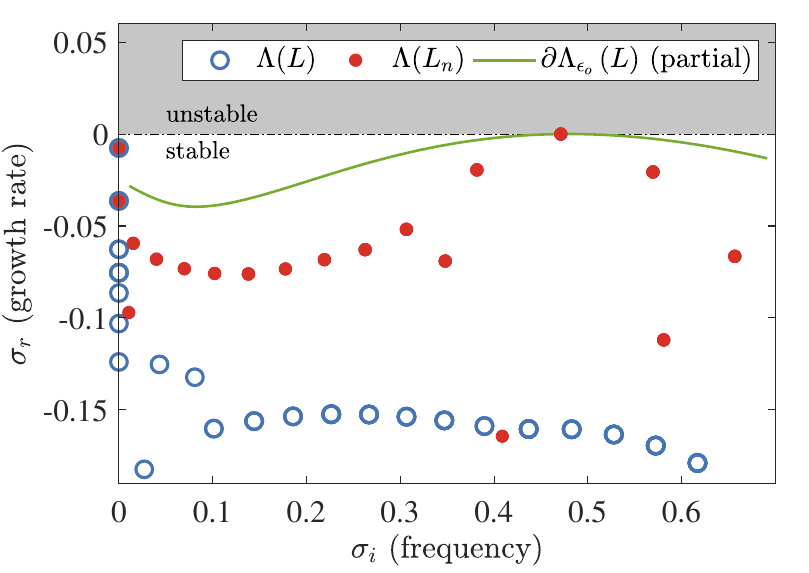}
\caption{\label{fig:opert} 
Natural and perturbed spectra of the flow past a backward-facing step (sketched in figure~\ref{fig:sbfs}a) at $Re=500$.
Blue circles:  eigenvalues of the linearised Navier-Stokes operator $L$. 
Red dots: eigenvalues of the  linear operator perturbed with  $\ez P  = \ez \hbf_o \left \langle  \hbu_o,*\right \rangle$. 
By construction, one eigenvalue of  $\hL_n = L + \ez P$  lies on the imaginary axis.
Green isocontour: part of the $\ez$-pseudospectrum of $L$, where $||R(z)||=1/\ez$. 
By construction, the $\ez$-pseudospectrum is contained in the stable half-plane, except at $i\wz$ where it touches the neutral axis.}
\end{figure}

As an illustration of the fact that a small-amplitude perturbation can easily ``neutralise'' a nonnormal operator, we consider the  Navier-Stokes operator linearised around the steady flow past a backward-facing step (BFS), sketched in figure~\ref{fig:sbfs}, at $Re=500$. 
The most amplified frequency $\wz = \wzm \approx 0.47$ is associated with  $\ez \approx 1.3 \cdot 10^{-4} \ll 1$. The spectra of $L$ and $\hL_n$ are shown in figure~\ref{fig:opert}, together with part of the $\ez$-pseudospectrum of $L$. 
Clearly, the very small perturbation  $\ez P$ locates an eigenvalue exactly onto $i\wz$, despite the strong stability of $L$. We stress that neither $\wz$ nor $\ez$ can be deduced only by inspecting the spectrum of $L$.

Nevertheless, in what follows, it is really the inverse resolvent and not the state operator $L$ that we propose to perturb. Indeed, $L$ is generally a real operator whereas $\hL_n$ is necessarily a complex one, and only one side of the spectrum of $\hL_n$ can generally be made neutral at a time, depending on whether $L$ is perturbed with $P$ or its complex conjugate $P^*$.

The inverse gain  $\ez \ll 1$ constitutes a natural choice of small parameter. We choose the Navier-Stokes equations for their nonlinear term $(\bm{U} \cdot \nabla) \bm{U}$, which yields both a nonnormal linearised operator and a rich diversity of behaviours.
The flow is \textit{weakly} forced  by $\bm{F} = \phi \sqrt{\ez}^{3} \bm{\hat{f}}_h e^{i \wz t} + c.c $, where $\bm{\hat{f}}_h$ is an arbitrary (not necessarily optimal) forcing structure, and $\phi=O(1)$ is a real  prefactor. 
Imposing  $ || \bm{\hat{f}}_h  || = 1$, the  forcing amplitude is $F \doteq \phi\sqrt{\ez}^{3}$. 
A separation of time scales is invoked for the flow response: its envelope is assumed to vary on a slow time scale $T = \ez t$ (such that $\mathrm{d}_t = \partial_t + \ez \partial_T$). This ensures a comprehensive distinguished scaling and suggests the following  multiple-scale expansion: 
\begin{align}
\bm{U}(t,T) = \bm{U}_e + \sqrt{\ez}\bm{u}_1(t,T)  + \ez \bm{u}_2(t,T) + \sqrt{\ez}^3 \bm{u}_3(t,T) + O(\ez^2).
\label{eq:ms}
\end{align}
The velocity field at each order $j$ is then Fourier-expanded as
\begin{equation}
\bu_j(t,T) = \obu_{j,0}(T) + \sum_m (\obu_{j,m}(T)e^{i m \wz t} +c.c),
\label{eq:res-fourier}
\end{equation}
\noindent $m =1, 2, 3 \ldots$. This decomposition is certainly justified in the permanent regime, of interest in this analysis. The proposed slow dynamics does not aim to capture the transient regime but flow variations around the permanent regime. Introducing (\ref{eq:ms})-(\ref{eq:res-fourier}) into the Navier-Stokes equations and using (\ref{eq:pinvr}) to perturb the operator $\Ri$ appearing from time derivation yields
\begin{align}
\begin{split}
& \sqrt{\ez} \Big[ \left ( \Phi \obu_{1,1}e^{i \wz t} + c.c \right) + \bs_1 \Big] +
 \ez \Big[ \left (  \Phi\obu_{2,1}e^{i \wz t} + c.c \right) + \bs_2 
+   C(\bu_1,\bu_1) \Big] + \\
& \sqrt{\ez}^3 \Big[ \left ( \Phi\obu_{3,1}e^{i \wz t} + c.c \right) + \bs_3  
+ 2 C(\bu_1,\bu_2) + \partial_T \bm{u}_1 + \left ( P \obu_{1,1}e^{i \wz t} + c.c \right) \Big] + O(\ez^2) \\ 
& = \phi \sqrt{\ez}^3 \bm{\hat{f}}_h e^{i \wz t} + c.c,
\end{split}
\end{align}
where 
\begin{equation*}
\bs_j \doteq -L\bm{\overline{u}}_{j,0}(T) + \left[\sum_{m}^{}(im\wz-L)\obu_{j,m}(T)e^{i m \wz t} + c.c \right].
\end{equation*}
For $m=2,3,...,$ and $C(\bm{a},\bm{b}) \doteq \frac{1}{2}((\bm{a} \cdot \nabla)\bm{b} + (\bm{b} \cdot \nabla) \bm{a})$.  Note that the perturbation $\ez P$ modifying $R(i\omega_0)^{-1}$ into $\Phi$ at leading order is compensated for at third order. Terms are then collected at each order in $\sqrt{\ez}$, leading to a cascade of linear problems, detailed hereafter. 

At order $\sqrt{\ez}$, we collect 
$(i m  \wz I  - L)  \obu_{1,m} = \bm{0} $ for $m=0,2,3 \ldots$, and  $\Phi \obu_{1,1} = \bm{0}$.  
Since $L$ is strictly stable, the unforced equation for $m \neq 1$ can only lead to $\obu_{1,m}=\bm{0}$. 
Conversely, the kernel of $\Phi$ contains the optimal response $\hbu_o$, therefore $\obu_{1,1}(T)=A(T)\hbu_o$, where $A(T) \in \mathbb{C}$ is a slowly-varying scalar amplitude verifying $\partial_t A=0$. 
Finally, the general solution at order $\sqrt{\ez}$ writes
\begin{equation}
\bm{u}_1(t,T) = A(T)\hbu_o e^{i\wz t} + c.c.
\label{eq:o1}
\end{equation}

At order $\ez$, we obtain the solution $\bm{u}_2 = \left | A \right |^2 \bm{u}_{2,0}  + \left(A^2 e ^{2i\wz t}\hbu_{2,2} + c.c \right)$, where
\begin{equation}
\begin{split}
 - L  \bm{u}_{2,0} &=  - 2 C(\hbu_o,\hbu^*_o),
 \\
(2 i   \wz I  - L)  \hbu_{2,2} &=  -  C(\hbu_o,\hbu_o).
\end{split}
\label{eq:lso2}
\end{equation}
The homogeneous solution of the system $\Phi \obu_{2,1} =  \bm{0}$ is arbitrarily proportional to $\hbu_o$, and written $A_2(T)\hbu_o$. It can be ignored ($\obu_{2,1} =  \bm{0}$) without loss of generality. As shown in \cite{Fujimura91}, it could also be kept, provided it is included in the definition of the amplitude, which would then become $A + \ez A_2$. 

At order $\sqrt{\ez}^3$, we assemble two equations yielding the Fourier components of the solution oscillating at $\wz$,  
\begin{equation}
\Phi \obu_{3,1} = - A|A|^2 \left[ 2C(\hbu_o,\bm{u}_{2,0}) + 2C(\hbu_o^*,\hbu_{2,2}) \right] -\hbu_o\frac{\mathrm{d} A}{\mathrm{d} T} - A \hbf_o + \phi \hbf_h
\label{eq:ro3}
\end{equation}
(recalling $P\hbu_o = \hbf_o$), and at $3\wz$, $(3 i   \wz I  - L)  \obu_{3,3} = 2 A^3 C(\hbu_o, \hbu_{2,2})$. The operator $\Phi$ being singular, the only way for $\obu_{3,1}$  to be non-diverging, and thus for the asymptotic expansion to make sense, is that the right-hand side of (\ref{eq:ro3}) has a null scalar product with the kernel of $\Phi^\da$,
i.e. is orthogonal to the adjoint mode $\hbf_o$ associated with $\hbu_o$. 
This is 
known as the ``Fredholm alternative''. As a result, the amplitude $A(T)$ satisfies 
\begin{equation}
\frac{1}{\eta}\frac{\mathrm{d}A}{\mathrm{d}T} =  \phi \gamma - A  - \frac{\mu + \nu}{\eta}A\left | A \right |^2,
\label{eq:wnlr}
\end{equation}
with the coefficients
\begin{align}
& \eta = \frac{1}{\left \langle  \hbf_o , \hbu_o \right \rangle  },\quad  
 \gamma =  \left \langle  \hbf_o , \hbf_h \right \rangle, \nonumber \\ 
& \frac{\mu}{\eta} = \left \langle  \hbf_o  , 2C(\hbu_o,\bu_{2,0}) \right \rangle, \quad 
\frac{\nu}{\eta} = \left \langle  \hbf_o  , 2C(\hbu^*_o,\hbu_{2,2}) \right \rangle. 
\label{eq:wnlcoeff}
\end{align}
The coefficient $\gamma$ is the projection of the applied forcing on the optimal forcing. The coefficient $\mu$ embeds the interaction between $\hbu_o$ and the static perturbation $\bu_{2,0}$, i.e. it corrects the gain according to the fact that $\hbu_o$ extracts energy from the \textit{time-averaged mean} flow rather than from the original \textit{base} flow.  
We show in  Appendix~\ref{appendix:S2} that, in the regime of small variations around the linear gain, the amplitude equation reduces to the standard sensitivity of the gain \cite{Brandt11} to a base flow modification induced by $\bu_{2,0}$. In contrast, the coefficient $\nu$ embeds the interaction between $\hbu_o^*$ and the second harmonic $\hbu_{2,2}$. 
Introducing the rescaled quantities  $a \doteq \sqrt{\ez} A$ and $F = \phi \sqrt{\ez}^3$, such that the weakly nonlinear harmonic gain $G = \left \| \sqrt{\ez} \obu_{1,1} \right \| / \left \|\phi \sqrt{\ez}^3 \hbf_h  \right \|$ is simply $=\left | a \right |/F$,  (\ref{eq:wnlr}) becomes

\begin{equation}
\frac{1}{\eta \ez }\frac{\mathrm{d} a}{\mathrm{d} t} =  \frac{ \gamma F}{\ez} - a - \frac{\mu + \nu}{\eta \ez } a \left | a \right |^2.
\label{eq:Ampeq}
\end{equation}
The gain associated with the linearised version of (\ref{eq:Ampeq}) is $G = \left |\gamma \right|/\ez$,  as expected for the linear prediction. We recover  $G = 1/\ez$ when the optimal forcing is applied ($\gamma=1$). We also note that this expression predicts  $G=0$ when $\gamma=0$, which merely indicates that the linear response is orthogonal to $\hbu_o$, without stating anything on the gains associated with sub-optimal forcings except that they should be at most $O(\ez^{-1/2})$, assuming a sufficiently large ``spectral'' gap in the singular-value decomposition of the resolvent operator. For the rest of the paper, we set $\gamma=1$. Expressing $a$ in terms of an amplitude $\ampa \in \mathbb{R}^+$ and a phase $\rho \in \mathbb{R}$ such that $a(t)=\left | a(t) \right |e^{i \rho(t)}$, the time-independent equilibrium solutions, or \textit{fixed points}, of equation (\ref{eq:Ampeq}), named $(\Heq,\Peq)$, solve: 
\begin{align}
\frac{F}{\ez}e^{-i \Peq} =   \Heq + \frac{\mu + \nu}{\eta \ez } \Heq^3,
\label{eq:eqsol}
\end{align}
Squaring and adding the real and imaginary parts of (\ref{eq:eqsol}) leads to a third-order polynomial for the equilibrium amplitude of (\ref{eq:Ampeq}):
\begin{align}
D Y^3 + 2 B Y^2 + Y  = \left( \frac{F}{\ez} \right)^2 \quad \text{with} \quad D=\frac{ \left | \mu + \nu \right |^2 }{\ez^2 \left | \eta \right |^2} > 0 \quad \text{and} \quad B = \Re\left[\frac{\mu + \nu}{\ez\eta}\right]
 \label{eq:ea}
\end{align}
and where $Y =  \Heq^2 > 0$. Let $p(Y) = D Y^3 + 2 B Y^2 + Y $ be the left-hand side of (\ref{eq:ea}). We further distinguish two cases: (i) if $B \geq 0$ , $p(Y)$ is increasing monotonously with $Y$ and can only cross the constant line $(F/\ez)^2$ once. We have in addition $p(Y) > Y$, thus the gain smaller than the linear prediction and monotonously decaying while increasing  $F$. 
Conversely, if (ii) $B < 0$, we have $p(Y) < Y$ in the interval $0<Y<-2B/D$, and the gain should then be greater than the linear one in the corresponding range of forcing $0 < (F/\ez)^2 < -2B/D$. Furthermore, $p(Y)$ may vary non-monotonously over this interval and cross the constant line $(F/\ez)^2$ three times (leading to three solutions for $Y$); namely, $p(Y)$ may be decreasing on a certain interval of $Y$ while dominated by the negative term $\propto Y^2$, bridging two other intervals where $p(Y)$ is increasing due to the respective positive terms $\propto Y$ and $\propto Y^3$. A necessary and sufficient condition for such a case to occur is that the equation $\mathrm{d}P/\mathrm{d}Y = 3D Y^2 + 4BY +1 = 0$ possesses two real and and positive solutions. This is guaranteed if and only if the determinant $\Delta \doteq 16B^2-12D$ is strictly positive. Finally, for $-2B/D \leq Y$, $p(Y)$ must be monotonously increasing again with $p(Y)\geq Y$, resulting in a gain smaller than the linear one and monotonously decreasing while increasing $F$.

The stability of the limit cycle associated with the equilibrium solution(s) $(\Heq,\Peq)$ can be established from the amplitude equation (\ref{eq:Ampeq}). Although in a different context, this was demonstrated for instance in \cite{Tuckerman90}, where the bifurcation diagram of the Eckhaus instability is determined directly from the Ginzburg-Landau equation for the envelope of the critical eigenfunction. Equation (\ref{eq:Ampeq}) can be expressed as a two-by-two amplitude/phase nonlinear dynamical system: 
\begin{align}
\frac{\mathrm{d} \ampa}{\mathrm{d} t} = F \left[\eta_r \cos(\rho) + \eta_i \sin(\rho)\right] - \eta_r \ez \ampa - (\mu_r+\nu_r) \ampa^3 \label{eq:nldy_amp} \\
\ampa \frac{\mathrm{d} \rho}{\mathrm{d} t} = F \left[\eta_i \cos(\rho)-\eta_r \sin(\rho)\right] - \eta_i \ez \ampa - (\mu_i+\nu_i) \ampa^3.
\label{eq:nldy_amp2}
\end{align}
Perturbing this system around the equilibrium solution $(\Heq,\Peq) + (\ampa'(t),\rho'(t))$ and neglecting nonlinear terms leads to the following equation for the perturbation $\mathrm{d}_t (\ampa',\rho')^T = J \cdot (\ampa',\rho')^T$ where $J$ is the Jacobian matrix expressed as 
\begin{equation}
J
 = \begin{bmatrix}
-\ez \eta_r - 3 (\mu_r+\nu_r) \Heq^2 & F\left[ \eta_i\cos(\Peq) - \eta_r\sin(\Peq) \right]\\ 
-\ez \eta_i\Heq^{-1} - 3 (\mu_i+\nu_i) \Heq & -F\left[ \eta_i\sin(\Peq) + \eta_r\cos(\Peq) \right]\Heq^{-1}
\end{bmatrix}.
\label{eq:Jacob}
\end{equation}
If at least one of the two eigenvalues of $J$ has a positive real part, the associated equilibrium is linearly unstable.

Note that equations (\ref{eq:nldy_amp}) and (\ref{eq:nldy_amp2}) for the amplitude and the phase of the oscillating linear response, are similar to those that would be obtained for a classical \textit{Duffing-Van der Pol} oscillator with appropriate parameters and harmonically forced around its natural frequency. If the latter is set to one, $\eta_r \ez$ and $\eta_i \ez$ are respectively proportional to the damping ratio and the detuning parameter. The coefficient $(\mu_i+\nu_i)$ is proportional to the the cubic stiffness parameter (\textit{Duffing} nonlinearity $\propto x^3$), and $(\mu_r+\nu_r)$ to the nonlinear damping parameter (\textit{Van der Pol} nonlinearity $\propto \dot{x}x^2$).

For the sake of completeness,  Appendix~\ref{appendix:Pois1} shows how to compute higher-order corrections of (\ref{eq:Ampeq}). It is worth mentioning, in particular, that the action of $\Phi$ need not be computed explicitly and can be replaced by the action of  $(i\wz I-L)$ for all practical purposes.

\subsection{Application case: the flow past a backward-facing step}
\label{sec:harmonicBFS}

Equation (\ref{eq:Ampeq}) is the first main result of this study and will be further referred to as the Weakly Nonlinear Nonnormal harmonic (WNNh) model. 
We discuss its performance when the stationary flow past a BFS sketched in figure~\ref{fig:sbfs} is forced harmonically with the optimal structure $\hbf_o$. At $Re=500$ and the optimal forcing frequency, $\hbf_o$ is shown in figure~\ref{fig:fobfs} together with is associated response $\hbu_o$ in figure~\ref{fig:uobfs} (see Appendix~\ref{appendix:S3} for details about the geometry and the numerical method).
\begin{figure}
\centering
  \begin{subfigure}[b]{0.65\linewidth}
\includegraphics[trim={1.8cm 0.05cm 1.2cm 0.1cm},clip,width=1\linewidth]{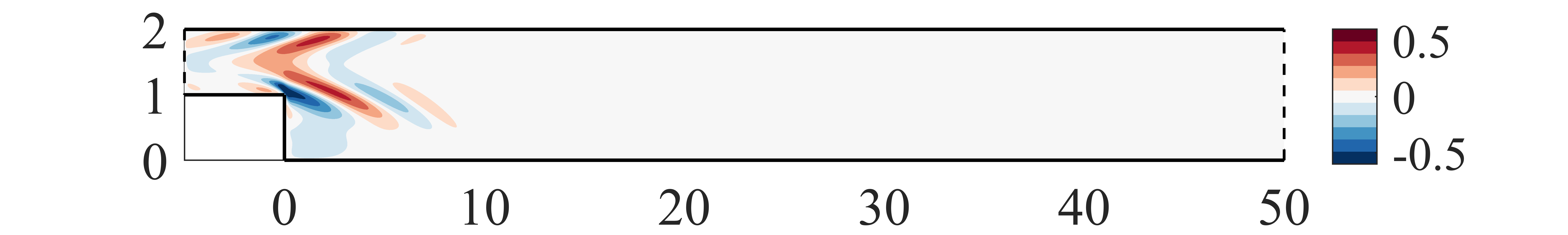}
\caption{\label{fig:fobfs}}
\end{subfigure}
  \hfill
  \begin{subfigure}[b]{0.65\linewidth}
\includegraphics[trim={1.8cm 0.05cm 1.2cm 0.1cm},clip,width=1\linewidth]{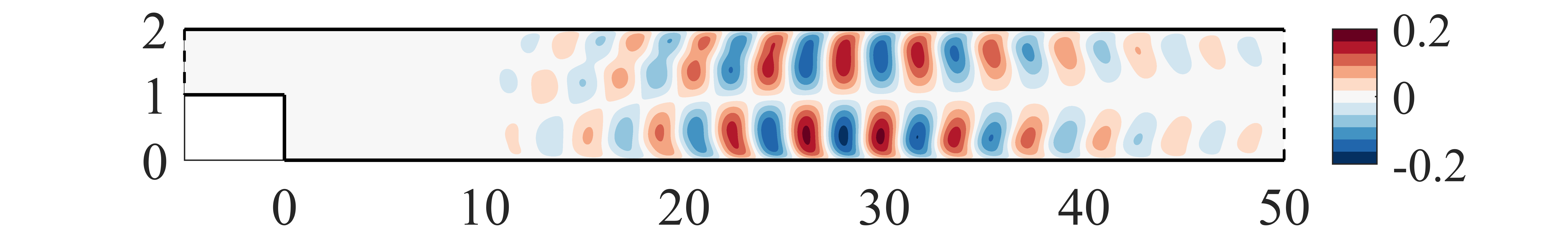}
\caption{\label{fig:uobfs}}
 \end{subfigure}
\caption{\textbf{(a)} Streamwise ($x$) component of the optimal harmonic forcing structure $\Re(\hbf_o)$ for the BFS (sketched in figure~\ref{fig:sbfs}a) at $Re=500$ and at the optimal forcing frequency $\wz/(2\pi) = \wzm/(2\pi) = 0.075$.  
\textbf{(b)}~Streamwise component of the associated response $\Re(\hbu_o)$. Both structures are normalised as $||\hbf_o|| = ||\hbu_o || = 1$.}
\end{figure}
As shown in \cite{Blackburn08,Boujo15}, the BFS flow constitutes a striking illustration of streamwise nonnormality. As seen in figure~\ref{fig:fobfs}, the optimal forcing structure is located upstream and triggers a spatially growing response along the shear layer adjoining the recirculation region, as the result of the convectively unstable nature of the shear layer. We first set the Reynolds number $Re$ between $200$ and $700$, and the frequency $\wz= 2 \pi \times  0.075$ close to the most linearly amplified frequency $\wzm$, which varies only slightly with $Re$.
The linear gain  grows exponentially with $Re$ \cite{Boujo15}, as seen in table \ref{tab:coeffBFS}. 
Since $\eta$ scales like $O(\ez^{-1/2})$, the term in $\mathrm{d}A/\mathrm{d}T$ in  (\ref{eq:ro3}) is asymptotically consistent only close to equilibrium points where $\mathrm{d}A/\mathrm{d}T = 0$, which is the regime of primary interest in the context of harmonic forcing. In accordance, the temporal derivative $\mathrm{d}A/\mathrm{d}T$ is kept in (\ref{eq:wnlr}) to assess the stability of such equilibria, determined by the analysis of the Jacobian matrix (\ref{eq:Jacob}).
\begin{table}
  \begin{center}
  \def~{\hphantom{0}}
\begin{tabular}{lccccc}
 $Re$ & $\epsilon_0$ & $\eta$ & $\mu/(\ez\eta)$ & $\nu/(\ez\eta)$ & $B$\\ \hline
 $200$ & $ 73.9^{-1} $ & $ 3.66 + i\cdot 0.0163 $ & $ 5.13 + i\cdot 1.32 $ & $0.137 - i\cdot 1.13$ & $5.27$\\
$500$ & $7456.6^{-1}$ & $117.1 + i \cdot 0.653 $ & $8.23 + i \cdot 2.60$ & $0.364 + i \cdot 0.396$ & $8.59$\\
$700$ & $148080^{-1}$ & $1626.7 + i \cdot 8.65 $ & $9.06+ i \cdot 4.38$ & $-0.729 + i \cdot 1.39$ & $8.33$\\
\end{tabular}
\caption{\label{tab:coeffBFS} WNNh coefficients for the backward-facing step flow, when the optimal forcing structure ($\gamma=1$) is applied at the optimal frequency $\wz/(2\pi) = \wzm/(2\pi) = 0.075$.}
\end{center}
\end{table}
\begin{figure} 
\centering
  \begin{subfigure}[b]{0.49\linewidth}
\includegraphics[trim={3.15cm 10cm 4cm 10.20cm},clip,width=1\linewidth]{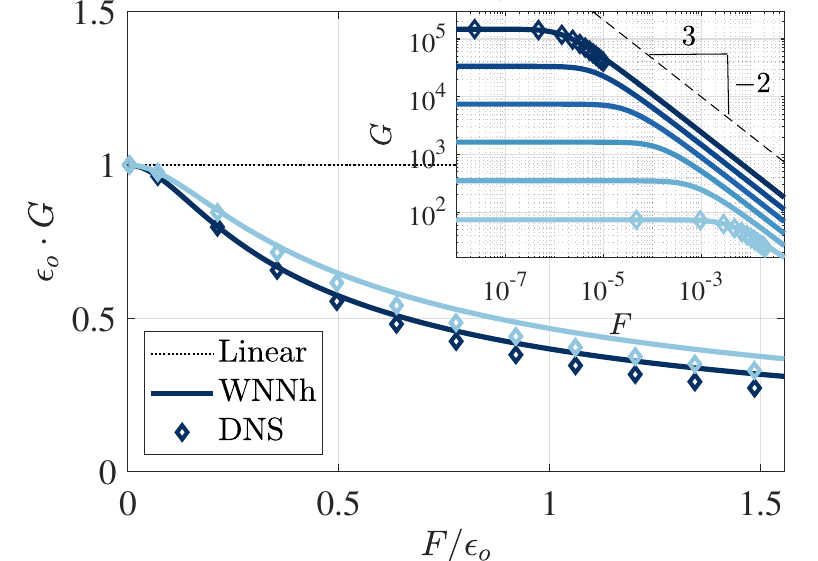}
\caption{\label{fig:bfs_a}}
\end{subfigure}
  \hfill
  \begin{subfigure}[b]{0.49\linewidth}
\includegraphics[trim={3.15cm 10cm 4cm 10.20cm},clip,width=1\linewidth]{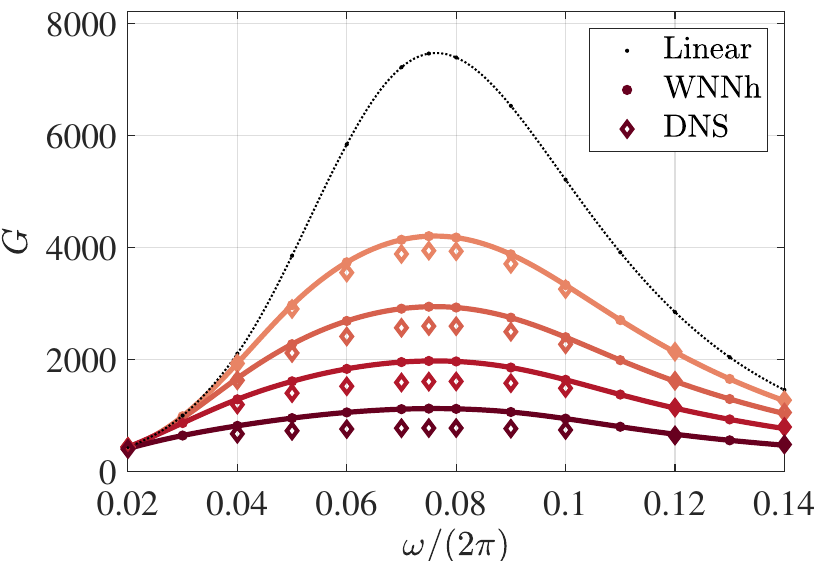}
\caption{\label{fig:bfs_b}}
 \end{subfigure}
\caption{\label{fig:ANL_BFS} 
Weakly and fully nonlinear harmonic gain in the backward-facing step flow  (sketched in figure~\ref{fig:sbfs}a). At each frequency and each Reynolds number, the optimal linear forcing structure $\hbf_o$ is applied.
\textbf{(a)}~Fixed frequency $\wz/(2\pi) = 0.075$,  varying Reynolds number $Re=200$ and $700$ (larger $Re$ darker).
Inset: log-log scale, $Re=200, 300 \ldots 700$. 
\textbf{(b)}~Fixed Reynolds number  $Re=500$, varying forcing $rms$ amplitude $F = \sqrt{2}^{-1} [1,2,4,10] \cdot 10^{-4}$ (larger amplitudes darker).} 
\end{figure}

Predictions from the WNNh model are compared to fully nonlinear gains extracted from direct numerical simulations (DNS) in figure~\ref{fig:bfs_a}. 
The DNS gains are the ratio between the temporal \textit{rms} of the kinetic energy of the fluctuations at $\wz$ (extracted through a Fourier transform) and the \textit{rms} of the kinetic energy of the forcing (for instance, the forcing $F \hbf_o e^{i \wz t} + c.c$ with $||\hbf_o||=1$ corresponds to an effective forcing \textit{rms} amplitude  of $\sqrt{2}F$). 
Since the coefficient $B$  defined in (\ref{eq:ea}) is strictly positive for all $Re$, the WNNh model predicts nonlinearities to saturate the energy of the response, and thus the gain to decrease monotonously with the forcing amplitude. This is confirmed by the comparison with DNS, displaying an excellent overall agreement.
As shown in the inset (in logarithmic scale), the nonlinear gain transitions from a constant value in the linear regime to a $-2/3$ power-law decay when nonlinearities prevail, as predicted from (\ref{eq:Ampeq}). 
This transition is delayed when the Reynolds number (and therefore the linear gain) decreases, and compares well with DNS data. 
The main plot (in linear scale) confirms the agreement with the DNS, and the improvement over the linear model. Re-scaled WNNh curves  appear similar for $Re=200$ and $Re=700$, and a slight overestimate is observed as the forcing amplitude approaches $\e_0$. 
Indeed, $F \sim \e_0$ implies $\phi \sim 1/\sqrt{\ez}$, which jeopardises the asymptotic hierarchy. 
Nonetheless, the error remains small for this flow  in the considered range of forcing amplitudes. 
Further physical insight is gained from the WNNh coefficients gathered in table \ref{tab:coeffBFS}.
The nonlinear coefficients remain of order one, which confirms the validity of the chosen scalings.  
The real part of $\mu$ being larger than that of $\nu$, the present analysis rationalises \textit{a priori} the predominance of the mean flow distortion over the second harmonic in the saturation mechanism reported \textit{a posteriori} in \cite{Lugo16}.

Next, we select $Re=500$ and report in figure~\ref{fig:bfs_b} harmonic gains as a function of the frequency, for increasing forcing amplitudes. 
At each frequency, the corresponding optimal forcing structure $\hbf_o$ is applied. 
The comparison between  DNS and  WNNh  is  conclusive over the whole range of frequencies. The saturating character of nonlinearities is well captured.
Such a good agreement may appear surprising in the low-frequency regime, for instance at $\wz / (2\pi) = 0.04$ where the second harmonics at frequency $2\omega_0$ could in principle be amplified approximately four times more than the fundamental. 
It  happens, however, that the associated forcing structure $-C(\hbu_o,\hbu_o)$ is located much farther downstream than the optimal forcing at $2\wz$, with a weak overlap region which results in a poor projection. 
Therefore, the second-order contribution does not reach amplitudes of concern in this flow, as a consequence of its streamwise nonnormality.

\subsection{Application case: Orr mechanism in the plane Poiseuille flow}
\label{sec:harmonicPoiseuille}

The weakly nonlinear evolution of the harmonic gain is now sought for the plane Poiseuille flow sketched in figure~\ref{fig:sbfs}, a typical flow with component-wise nonnormality \cite{Trefethen93,Schmid07}. Periodicity is imposed in the streamwise and spanwise directions with wavenumbers $k_x$ and  $k_z$, respectively. The set of parameters $(Re,k_x,k_z) = (3000,1.2,0)$ is selected. According to the classical work of \cite{Orzag71}, the base flow at this $Re$ number is linearly stable since instability first occurs at $Re_{cr} \approx 5772$ and $k_{x,cr} \approx 1.02$. In both the linear and nonlinear computations, the spanwise invariance $k_z=0$ is systematically maintained. While the base flow $U(y)$ has only one velocity component and depends only on one coordinate, the perturbations are here two-dimensional (i.e, $\bu = (u_x(x,y),u_y(x,y))$). The computations are performed in the streamwise-periodic box $(x,y) \in [0,2\pi/k_x] \times [-1,1] \equiv  \Omega$. All the scalar products are taken upon integration inside this periodic box, in particular for the normalisation $\langle \hbu_o,\hbu_o\rangle = \langle \hbf_o,\hbf_o\rangle =1$, and latter for the evaluation of the weakly nonlinear coefficients.

\begin{figure}
\centering
\includegraphics[trim={3cm 8cm 4cm 7.75cm},clip,width=0.6\linewidth]{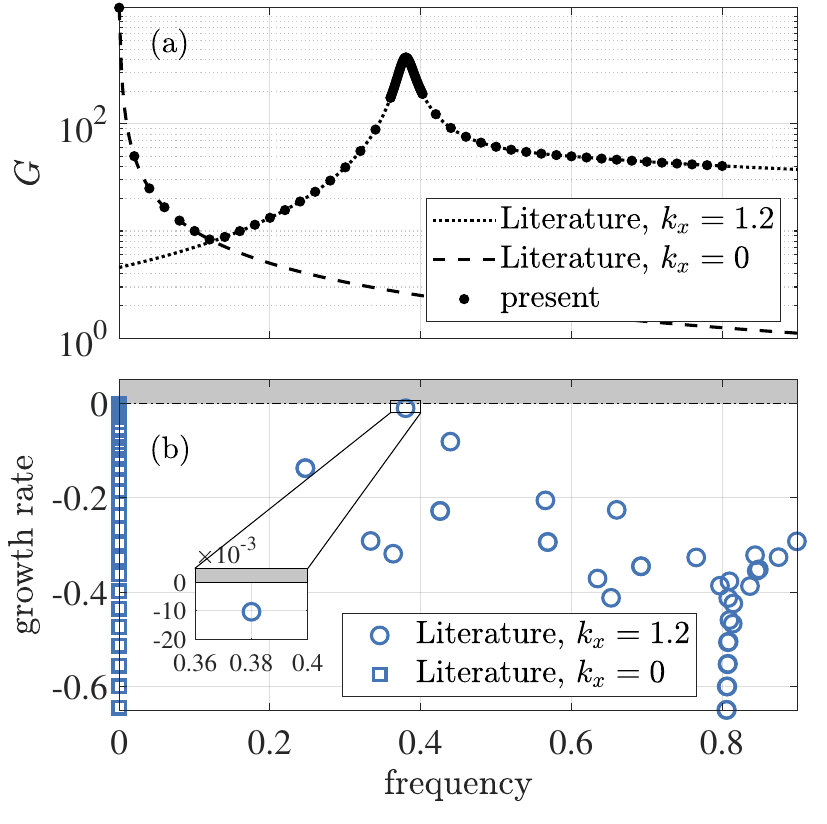}
\caption{\textbf{(a)} Linear harmonic (optimal)  gain as function of the optimisation frequency. Present results are compared to those reproduced from \cite{SH01}, where perturbations are expressed as Fourier mode of streamwise wavenumber $k_x$.
 \textbf{(b)}~Eigenspectra. 
 \label{fig:linpois}}
\end{figure}

\begin{figure}
\centering
   \begin{subfigure}[b]{0.49\linewidth}
\includegraphics[trim={0.1cm 2.5cm 1cm 3.5cm},clip,width=1\linewidth]{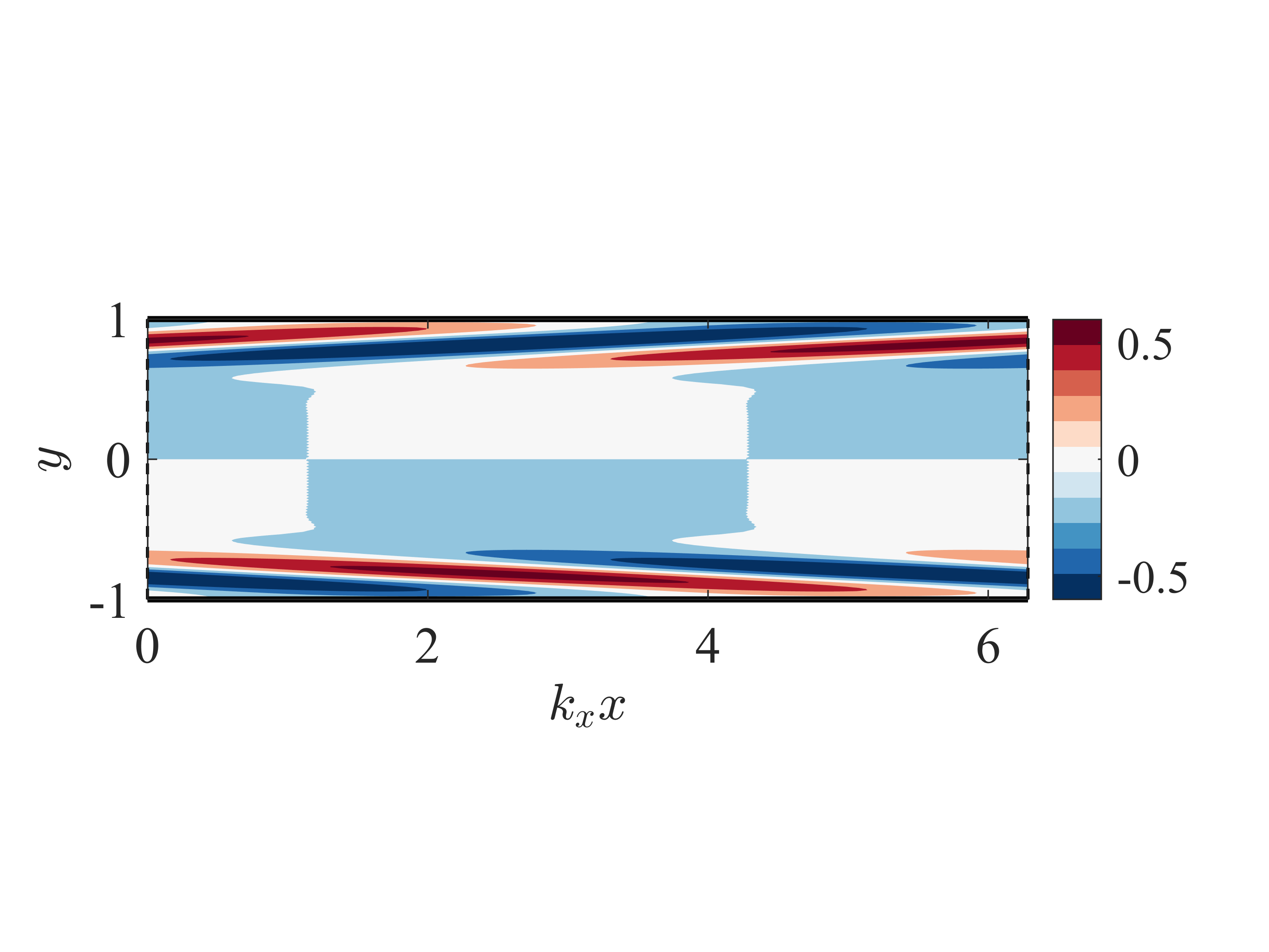}
\caption{\label{fig:fopoisr}}
\end{subfigure}
  \hfill
    \begin{subfigure}[b]{0.49\linewidth}
\includegraphics[trim={0.1cm 2.5cm 1cm 3.5cm},clip,width=1\linewidth]{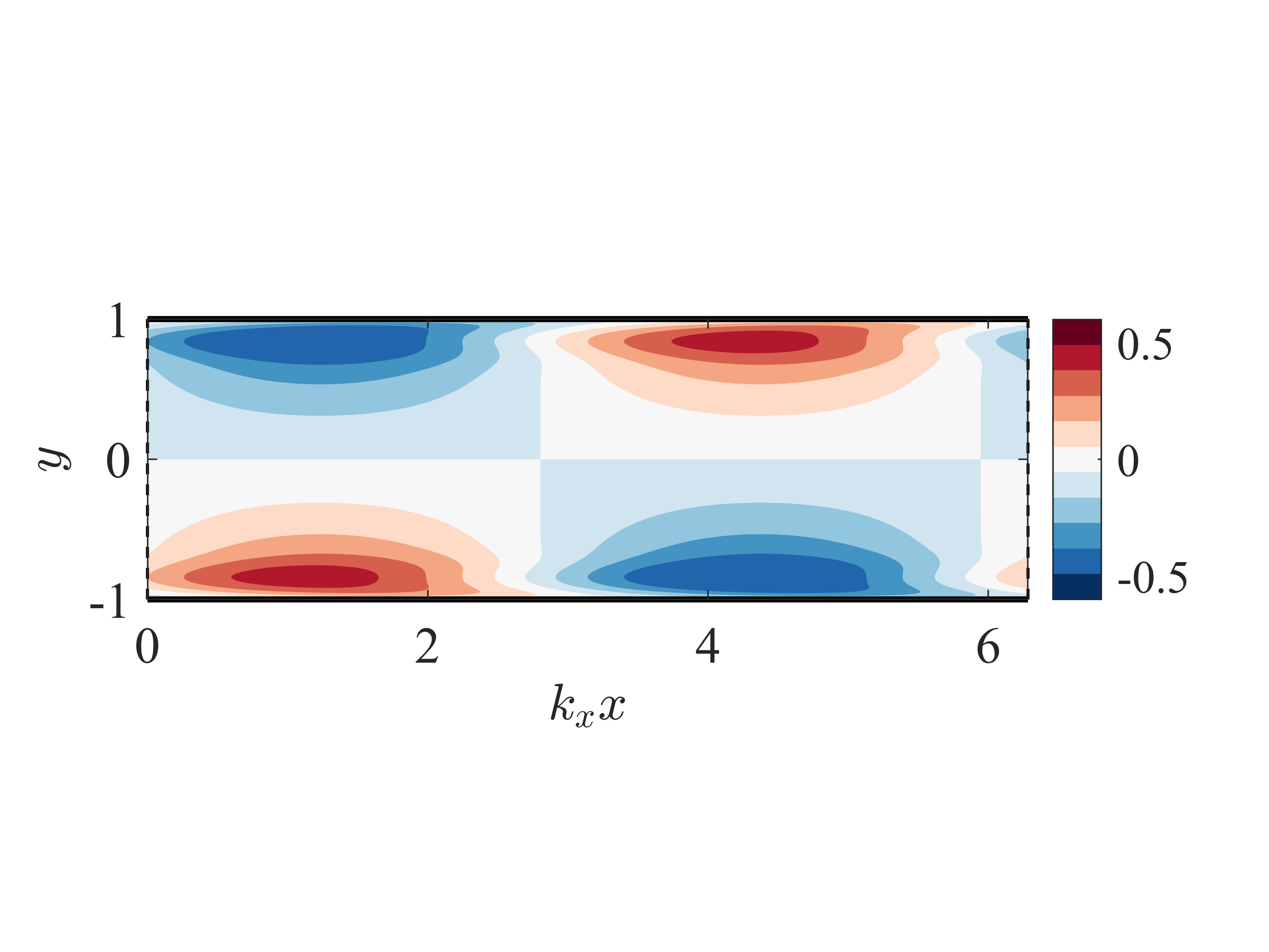}
\caption{\label{fig:uopoisr}}
\end{subfigure}
\caption{\textbf{(a)}~Streamwise component of the optimal forcing $\Re(f_{o,x})$ for the plane Poiseuille flow (sketched in figure~\ref{fig:spois}b) for $(Re,k_x,k_z) = (3000,1.2,0)$ and $\wz=0.3810$. 
\textbf{(b)}~Streamwise component of the response $\Re(\hat{u}_{o,x})$. Both fields are normalised as  $||\hbf_o|| = ||\hbu_o|| = 1$. Only one wavelength  $0 \leq k_x x \leq 2\pi$ is shown.}
\end{figure}

\begin{figure*} 
\centering
\begin{subfigure}[b]{0.49\linewidth}\includegraphics[trim={3.25cm 10cm 4cm 10.20cm},clip,width=1\linewidth]{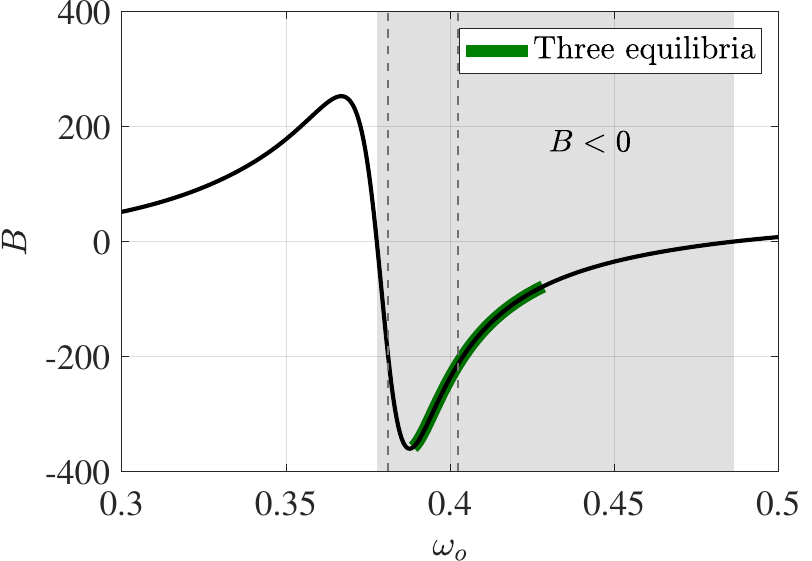}
\caption{\label{fig:wtfpois_a}}
\end{subfigure}
  \hfill
\begin{subfigure}[b]{0.49\linewidth}\includegraphics[trim={3.5cm 10cm 4cm 10.20cm},clip,width=1\linewidth]{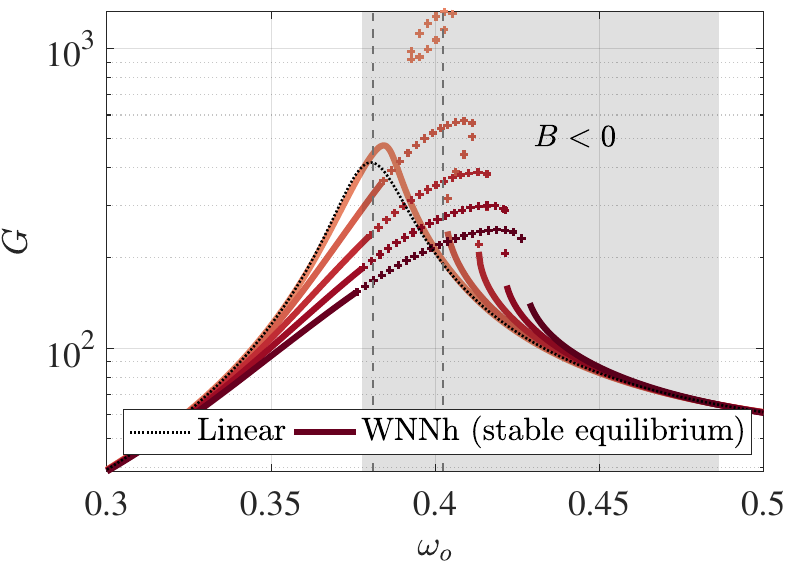}
\caption{\label{fig:wtfpois_b}}
 \end{subfigure}
\caption{\textbf{(a)} Coefficient $B$ defined in (\ref{eq:ea}) as a function of the optimisation frequency. The superimposed bold green line indicates that $B$ and $D$ are such that three equilibrium solutions to (\ref{eq:Ampeq}) exist. \textbf{(b)} Weakly nonlinear harmonic gain predicted by the WNNh model for increasing forcing amplitude $F$ in $[0.55, 1.45, 2.35, 3.25, 4.15] \cdot 10^{-4}$ (larger $F$ darker). Solid lines denote stable equilibrium solutions of (\ref{eq:Ampeq}) whereas bold plus markers (+) denote the unstable ones. The vertical dashed grey lines highlight $\wz=0.3810$ and $\wz=0.4025$. The grey zone denotes a negative $B$.
\label{fig:wtfpois}}
\end{figure*}

The linear optimal gain (\ref{eq:hgain}) is computed in the frequency interval $0 \leq \wz \leq 0.8$ (figure~\ref{fig:linpois}a), together with the associated optimal forcing and responses structures. Results are validated with the $1D$ results of \cite{SH01} based on a Fourier expansion of wavenumbers $k_x=1.2$ and $k_x=0$ in the streamwise direction. 
Eigenspectra are also reported in figure~\ref{fig:linpois}(b).
The SVD algorithm applied to the periodic box automatically selects the most amplified wavenumber among all spatial harmonics $n k_x$ with $n=0,1,2,..$. Below $\wz \approx 0.12$, the harmonic $0 \cdot k_x=0$ is dominant due to the concentration of weakly damped eigenvalues along the imaginary axis. The gain $G(\wz = 0) = 1216 $ is equal to the inverse of the smallest damping rate among all these spatially invariant modes. The large value of the gain associated with those modes is understood considering that the small pressure gradient $(2/Re,0)^T=(2/3000,0)^T$ is sufficient to induce the Poiseuille base flow (equal to unity in the centerline). 
Above $\wz \approx 0.12$, the fundamental wavenumber $1 \cdot k_x=1.2$ prevails.
The corresponding harmonic gain presents a local and selective maximum for $\wz = 0.38$, certainly linked to the presence of the weakly damped eigenvalue $\sigma_1 = -0.0103+0.380i$. 
Nevertheless, $G(\wz =0.38) = 416$ is significantly bigger than $1/0.0103 \approx 97$. This is a direct consequence of the nonnormality of the plane Poiseuille flow. Unlike the backward-facing step flow, the nonnormality at play here is not due to the presence of a convectively unstable region but to the Orr mechanism suggested for the first time in \cite{Orr07}. Namely, an initial condition or forcing field constituted of spanwise vortices tilted towards the upstream direction (fig.~\ref{fig:fopoisr}), tilts downstream under the action of the mean shear (fig.~\ref{fig:uopoisr}), which leads to a significant gain in the kinetic energy of the perturbation.
The coefficient $B$ is shown in figure \ref{fig:wtfpois_a}, and the associated WNNh prolongation of the harmonic gain in figure \ref{fig:wtfpois_b}. 
The coefficient $B$ is negative in the interval $0.378 \leq \wz \leq 0.486$, and $B$ and $A$ are such that three equilibrium amplitudes $\Heq$ exist for some values of $F$ in the sub-interval $0.389 \leq \wz \leq 0.428$. Among them, \textit{none} or \textit{only one} is found to be stable. Consequently, as the forcing amplitude is increased, the harmonic gain curve leans toward the higher frequencies in figure \ref{fig:wtfpois_b}; in the meantime, a frequency interval where no stable solution is predicted appears and grows larger.

Note that in the absence of a stable equilibrium, it is natural to consider completing (\ref{eq:Ampeq}) up to $O(\sqrt{\ez}^{5})$. 
It is shown in Appendix~\ref{appendix:Pois1}, however, that such an approach is problematic in the present case, because the non-oscillating forcing terms appearing at $O(\ez^2)$ excite the largely amplified static modes visible in figure \ref{fig:linpois} for $\wz =0$. The associated gains being  of order $1/\ez$, the mean flow correction terms at order $O(\ez^2)$ break the asymptotic hierarchy. This problem is not encountered at order $\ord(\ez)$, because the forcing $-2C(\hbu_o,\hbu_o^*)$ in (\ref{eq:lso2}) projects poorly on the optimal one for $\wz=0$, and $||\bu_{2,0}||$ remains of order unity. \newline

\begin{figure*} 
\centering
\begin{subfigure}[b]{0.49\linewidth}\includegraphics[trim={3.15cm 10cm 3.25cm 9.8cm},clip,width=1\linewidth]{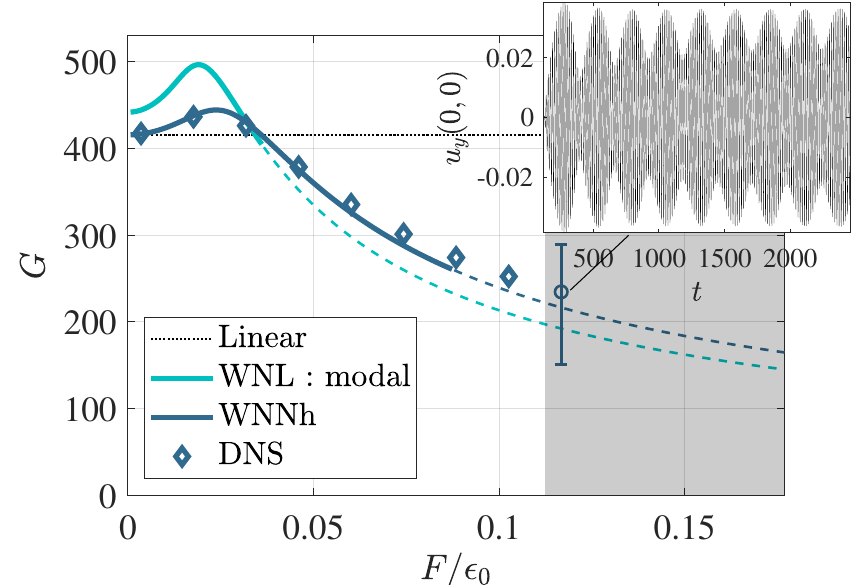}
\caption{\label{fig:ANL_Pois_a}}
\end{subfigure}
  \hfill
\begin{subfigure}[b]{0.49\linewidth}\includegraphics[trim={3.15cm 10cm 3.25cm 9.8cm},clip,width=1\linewidth]{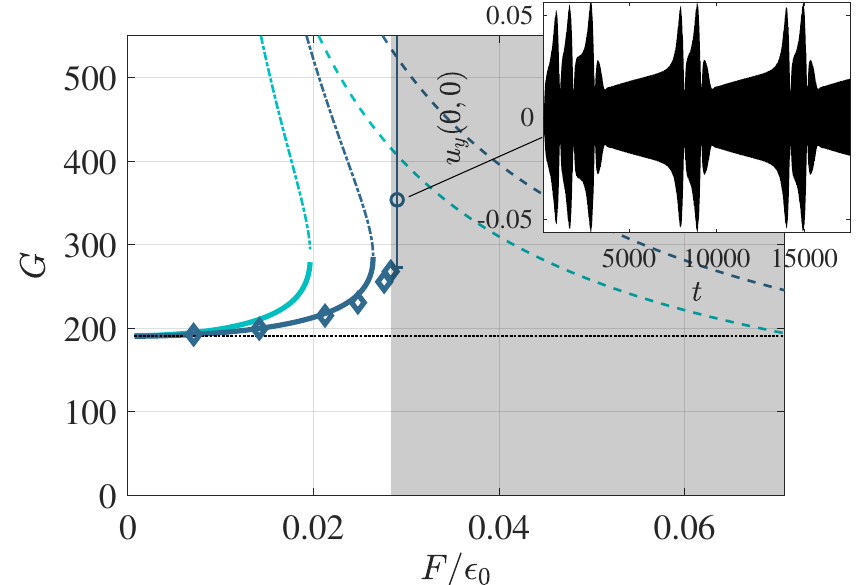}
\caption{\label{fig:ANL_Pois_b}}
\end{subfigure}
\caption{\label{fig:ANL_Pois} Evolution of the harmonic gain $G$ with respect to $F$ for \textbf{(a)} $\wz = 0.3810$ and  \textbf{(b)}~$\wz = 0.4025$. In both, the grey zone indicates that no harmonic gain could be properly defined, as the kinetic energy of the perturbation cease to converge to a constant value. In particular, the inset shows the monitoring of $u_y(0,0)$ for the flow represented by the circle (the link is indicated by a thin line).}
\end{figure*}

For comparison with DNS data, two different forcing frequencies with \textit{a priori} distinct behaviours are selected: $\wz=0.3810$ and $\wz=0.4025$. These two frequencies are highlighted by the vertical dashed grey lines in figure \ref{fig:wtfpois}. In both cases the coefficient $B$ is negative, and  for the case $\wz=0.4025$ three equilibrium solutions exist for some values of $F$. The linear gains and weakly nonlinear coefficients are reported in table \ref{tab:coeffPois}. 
\begin{table}
  \begin{center}
  \def~{\hphantom{0}}
\begin{tabular}{lccccc}
 $\wz$ & $\ez$ & $\eta$ & $\mu/(\ez \eta)$ & $\nu/(\ez \eta)$ & $B$   \\ \hline
 $0.3810$ & $ (415.6)^{-1} $ & $4.06 - i \cdot 0.218$ &$-177.0 + i \cdot 315.6$ & $ -17.4 + i \cdot 197.8i $ & $-194.4$    \\
$0.4025$ & $(190.4)^{-1}$ & $2.78 - i \cdot 3.89$ & $-160.1 - i \cdot 24.3 $ & $-52.3+ i \cdot 10.6$ &  $-212.4$ \\
\end{tabular}
\caption{WNNh coefficients for the plane Poiseuille flow at $(Re,k_x,k_z)=(3000,1.2,0)$, and when the optimal forcing structure ($\gamma=1$) is applied.
\label{tab:coeffPois}}
\end{center}
\end{table}
The corresponding WNNh prolongation of the linear gain as a function of the forcing amplitude is shown in figure \ref{fig:ANL_Pois}, together with DNS results. 
For comparison, the prediction of a ''classical'' (modal) amplitude equation constructed around the weakly damped eigenvalue $\sigma_1$ and its associated direct and adjoint modes is also added. Its derivation is briefly recalled in Appendix~\ref{appendix:Pois2}. 

For $\wz = 0.3810$ (figure \ref{fig:ANL_Pois_a}),  the WNNh gain initially increases with $F$ due to the negativity of $B$. As visible in table \ref{tab:coeffPois}, this is mostly due to the contribution of $\Re[\mu/(\ez \eta)]$ which is ten times larger than that of $\Re[\nu/(\ez \eta)]$. Thus, at this frequency, the principal factor for the initial increase of the WNNh gain is the Reynolds stress of the response $a \hbu_o$. 
The latter creates a mean flow that amplifies  the linear forcing $\hbf_o$ \textit{more} than the base flow does. This may be interpreted considering the displacement of the eigenvalue $\sigma_1$.  Let $\hbq_1$ (resp. $\hba_1$) denote the eigenmode (resp. adjoint mode) associated with the eigenvalue $\sigma_1$. The sensibility of the latter to the base flow deformation $\delta \bm{U}_b$ due to the Reynolds stress of $a \hbu_o$ writes:
\begin{equation}
     \delta \sigma_1 = -\frac{\left \langle \hba_1, C[\hbq_1,\delta \bm{U}_b] \right \rangle}{\left \langle \hba_1, \hbq_1 \right \rangle}
\end{equation}
where $\delta \bm{U}_b = |a|^2 \bu_{2,0}$. 
For $\wz = 0.3810$, we obtain $\delta \sigma_1 = |a|^2 ( 1.2 + i \cdot 3.9)$. Since $\Re[\delta \sigma_1]>0$, the eigenvalue is moving towards the unstable part of the complex plane under the action of the Reynolds stress. This is in accordance with the fact that the plane Poiseuille flow is subcritical, and may explain the initial increase in the gain with $F$. Meanwhile, $\Im[\delta \sigma_1]>0$ and $\sigma_1$ is shifting toward higher frequencies. Thus $\wz$ ceases to be the least damped frequency, which could shed light on the fact that increasing $F$ further leads to a monotonous decay in the WNNh gain at $\wz$. Because of the flow nonnormality, however, this explanation based solely on the location of $\sigma_1$  remains qualitative.

The overall agreement with the DNS results is excellent. Nevertheless, the WNNh model slightly underestimates the threshold in $F$ above which a stable equilibrium does not exist any more. It stands at $F/\ez = 0.087$ against $F/\ez =0.11$ for the DNS. This loss of a proper harmonic response may be symptomatic of the fact that $\sigma_1$ eventually crosses the neutral line and becomes unstable. Indeed for $F/\ez =0.11$ (blue circle in the grey zone in figure \ref{fig:ANL_Pois_a}), the FFT of the flow in its stationary regime presents two dominant neighbouring frequencies: the forcing one at $\omega = \wz$ and a second ``natural'' one at $\omega \approx 0.404$. As these two frequencies are very close, a beating behaviour is visible in the inset of figure \ref{fig:ANL_Pois_a} at a frequency consistent with $\Delta\omega=0.023$.

The classical modal amplitude equation leads to a prediction that is only qualitative. Even for $F=0$ the \textit{linear} harmonic gain  $|\langle \hbf_o , \hba_1 \rangle /(\langle \hbq_1 , \hba_1 \rangle \sigma_{1,r})|$ (see Appendix~\ref{appendix:Pois2} for its derivation) is overestimated, as it is deduced from the modal quantities linked to $\sigma_1$ \textit{only}. 
As mentioned earlier, in nonnormal flows a high number of eigenmode is generally necessary to describe its harmonic response, even in the presence of a weakly damped eigenvalue. Thus, relying on a single mode constitutes a poor description of the response to forcing.

We now consider $\wz=0.4025$, and the associated results in figure \ref{fig:ANL_Pois_b}. The WNNh model yields multiple equilibrium solutions in the range $0 < F/\ez < 0.0264$. Only the one represented by a thick continuous line is stable, and corresponds to a monotonous growth of the gain with $F$. The DNS results validate the existence of this solution. The two other solutions, depicted by the dash-dotted and dashed lines, are unstable in  one eigendirection and two eigendirections, respectively. Above $F/\ez = 0.0264$ the WNNh models predicts the loss of the stable equilibrium solution, which is accurately confirmed by the DNS whose threshold is located around $F/\ez = 0.0286$. Slightly above, the signal of $u_y(0,0)$ in the inset suggests again the presence of a ``natural'' frequency due to the subcritical destabilisation of $\sigma_1$. Indeed, $u_y(0,0)$ alternates between an algebraic growth typical of a true resonance (both natural and forcing frequencies collapse), and a beating-like behaviour whose period is very long (the natural frequency drifts slightly from the forcing one). 

Across this threshold, the evolution of the average kinetic energy of the response appears discontinuous. This \textit{loss} of a stable equilibrium is to be distinguished with its \textit{destabilisation} encountered for $\omega=0.3810$. Overall, the difference of behaviours between figures \ref{fig:ANL_Pois_a} and \ref{fig:ANL_Pois_b} may be explained by the difference of proximity between $\wz$ and $\Im[\sigma_1]$ of the mean flow. As the forcing is progressively increased above $F/\ez = 0.0286$, the flow response quickly becomes chaotic, and  then turbulent.

It should be mentioned that, in some situations, the amplitude equation (\ref{eq:Ampeq}) may be in default. 
First, as just mentioned, when the optimal linear harmonic gain at frequency $2\wz$ is $ \sim 1/\sqrt{\ez}$ or larger and projects well onto the optimal forcing, the asymptotic hierarchy is threatened as  $\hbu_{2,2}$ may be substantial enough to reach order $\sqrt{\ez}$ or above. 
It is thus important to assess that the norm of $\hbu_{2,2}$ remains of order one. 
A second delicate situation arises, for the same reason, when a sub-optimal gain at the frequency $\wz$ is $\sim 1/\ez$. 
In both cases, the model could be extended by including in the kernel of $\Phi$ the optimal response at frequency $2\wz$, or the sub-optimal response at frequency $\wz$, respectively. 

\section{Transient Growth}
\label{section:tg}

Next, we derive an amplitude equation for the weakly nonlinear transient growth in an unforced ($\bff=\bm{0}$) system, without restriction on its linear stability. 
The solution to the linearised equation (\ref{eq:ieqlin})  is $\bu(t) = e^{L t}\bu(0)$, where $e^{L t}$ is the operator exponential of $L t$. 
In an unforced context, the propagator $e^{ L t}$ maps an initial structure at time $t=0$ onto its evolution at $t \geq 0$. The largest linear amplification at $t_o>0$ (subscript $o$ for ``optimal'')  is
\begin{equation}
\centering
G(t_o)  =  \max_{\bu(0)} \frac{\left \|  \bu(t_o) \right \| }{\left \| \bu(0)\right \| } = \left \| e^{L t_o} \right \|  \doteq  \frac{1}{\ez}.
\label{eq:tgg}
\end{equation}
The singular value decomposition of the propagator $e^{L t_o}$ provides the transient gain $G(t_o)$ as the largest singular value of $e^{L t_o}$, as well as the left and right singular pair $\bv_o$ and $\bu_o$, respectively,
\begin{equation} 
e^{- L t_o} \bv_o = \ez \bu_o, \quad
\left[ \left( e^{ L t_o} \right)^\da \right]^{-1} \bu_o = \ez \bv_o,
\label{eq:1tg}
\end{equation}
where $||\bv_o|| = || \bu_o|| = 1$. The field $\bu_o$ is the optimal initial structure for the propagation time  $t=t_o$, and $\bv_o$ is its normalised evolution at $t_o$. The corresponding amplification is  $1/\ez$, as defined in (\ref{eq:tgg}). Smaller singular values are \textit{sub-optimal} gains, associated with orthogonal sub-optimal initial conditions. Their orthogonality is ensured by the fact that singular vectors of the operator $e^{L t_o}$ also are the eigenvectors of the symmetric operator $(e^{L t_o})^\da e^{L t_o}$, the singular values of the former being the square root of the eigenvalues of the latter. Of all the $t_o$, the time leading to the largest optimal gain will be highlighted with the subscript $m$ (for ``maximum'') such that $\max_{t_o>0} G(t_o) = G(t_{o,m})$.

By construction, the linear gain is  independent of the amplitude of the initial condition $\bu(0)$. 
As this amplitude increases, however, nonlinearities may come into play and the nonlinear gain may depart from the linear gain $G$.
Similar to the previous section on harmonic gain, we propose a method for capturing the effect of weak nonlinearities on the transient gain.

Due to the assumed nonnormality of $L$, the inverse gain is small, $\ez \ll 1$. 
While the previous section focused on the \textit{inverse resolvent}, it is now  the \textit{inverse propagator}  $e^{-Lt_o}$ that appears close to singular. The first equality of (\ref{eq:1tg}) can be rewritten as $(e^{- L t_o} - \ez \bu_o \langle \bv_o,*\rangle) \bv_o = \bm{0}$, which shows that the operator $(e^{- L t_o} - \ez \bu_o \langle \bv_o,*\rangle )$ is singular since $\bv_o \neq \bm{0}$ belongs to its kernel. 
Mirroring our previous reasoning for the WNNh model, we now wish to construct a perturbed inverse propagator whose kernel is the linear \textit{trajectory} 
\begin{equation}
\bl(t) \doteq \ez e^{L t} \bu_o 
\end{equation}
seeded by the optimal initial condition $\bu_o$ and normalised in $t=t_o$ such that $\bl(t_o) = \bv_o$.
One conceptual difficulty lies in that the linear response is not 
a fixed vector field, but a time-dependent trajectory;  therefore, the perturbed inverse propagator too should  depend on time. 
We propose to perturb the inverse propagator for all $t \geq 0$ as
\begin{equation}
\Phi(t) = e^{-L t} - \e_o P(t),
\quad \text{where} \quad 
P(t) \doteq  H(t) \frac{\bu_o \langle \bl(t),*\rangle}{\left \| \bl(t) \right \|^2},
\label{eq:pertP}
\end{equation}
and where the Heaviside distribution $H(t)$ satisfies $H(0)=0$ and $H(t>0)=1$. As the time $t \rightarrow t_o$, the perturbation operator $P \rightarrow \bu_o \langle \bv_o,*\rangle$ such that $||P|| \rightarrow 1$ and the expansion (\ref{eq:pertP}) is certainly justified. The non-trivial kernel of $\Phi(t)$ is $\bl(t)$ for all $t>0 $; the kernel reduces to $\bm{0}$ at $t=0$ since $\Phi(0)= I$. We show in addition that, for $t>0$, the non-trivial kernel of the adjoint operator $\Phi(t)^\da$ is
\begin{equation} 
\begin{split}
\ba(t) &\doteq \left( e^{L t} \right)^\da\bl(t).
\end{split}
\label{eq:adj}
\end{equation}
Indeed, using that $P^\da = \bl(t)\langle \bu_o,* \rangle/\left \langle \bl(t) , \bl(t) \right \rangle$ for $t>0$, we have
\begin{equation*}
\begin{split}
\Phi(t)^\da \ba(t) &= \left(e^{- L t}\right)^\da \ba(t) -  \ez \bl(t)  \frac{\left \langle \bu_o , \ba(t) \right \rangle}{\left \langle \bl(t) , \bl(t) \right \rangle}\\
& = \left(e^{- L t}\right)^\da \ba(t) -  \ez \bl(t)  \frac{\left \langle e^{L t}\bu_o , \bl(t) \right \rangle}{\left \langle \bl(t) , \bl(t) \right \rangle} \\
& = \left(e^{- L t}\right)^\da \ba(t) -  \bl(t) \\
& = \left[\left(e^{ L t}\right)^\da \right]^{-1} \ba(t) -  \bl(t) \\
& = \bm{0}.
\end{split}
\end{equation*}
\begin{figure}
\centering
\includegraphics[trim={5.5cm 10cm 5.5cm 10cm},clip,width=0.55\linewidth]{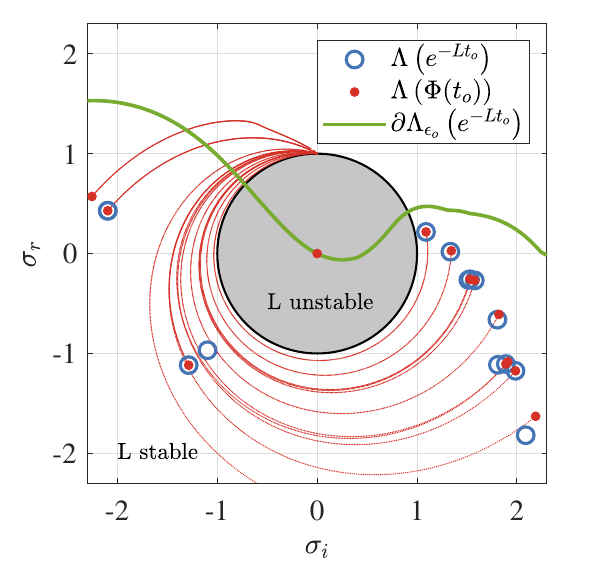}
\caption{
Restricted spectra (fifteen least 
stable eigenvalues) of the natural and perturbed inverse propagators of the plane Poiseuille flow (sketched in figure~\ref{fig:spois}b) 
for $t=t_o=10$ and $(Re,k_x,k_z)=(3000,0.5,2)$ (purely $1D$ computations  using the code of \cite{SH01} based on a Fourier expansion of wavenumbers $k_x$ and $k_z$ in $x$ and $z$, respectively). 
Blue circles: eigenvalues of $e^{-Lt_o}$.
Red dots: eigenvalues of $\Phi(t_o)$. By construction, one eigenvalue of $\Phi(t)$ lies at the origin. Thin red lines: full locus of the eigenvalues of $\Phi(t)$ for $t \leq t_o$. 
Green line: $\ez$-pseudospectrum of $e^{-Lt_o}$, such that $\left \| (e^{-Lt_o} -z I)^{-1} \right \| = 1/\ez$.
\label{eq:tgspec}}
\end{figure}

As an illustration of the singularisation of $e^{-Lt_o}$, parts of the spectra of  $e^{-Lt_o}$ and $\Phi(t_o)$ are shown in figure~\ref{eq:tgspec} for the plane Poiseuille flow sketched in figure~\ref{fig:spois}. The red dot at the origin is the null singular eigenvalue of $\Phi(t_o)$ associated with $\bl(t_o)$. Since $||P(t_o)||=1$, this singular eigenvalue lies on the $\ez$-pseudospectrum of $e^{-Lt_o}$, meaning that a perturbation of amplitude $\ez$ is sufficient to make the inverse propagator singular.

Recalling that $L$ is assumed strongly nonnormal, we choose $\ez \ll 1$ as expansion parameter, introduce the slow time scale $T = \ez t$, and propose the multiple-scale expansion
\begin{equation}
\bm{U}(t,T) = \bm{U}_e + \ez\bm{u}_1(t,T)  + \ez^2 \bm{u}_2(t,T) + O(\ez^3).
\label{eq:mstg}
\end{equation}
The square root scaling of the previous section is not made here, as resonance at second order cannot be excluded  \textit{a priori}. The flow is initialised with $\bm{U}(0) = \alpha \ez^2 \bu_o$, where $\alpha = O(1)$ is a prefactor.
After injecting this expansion in the unforced Navier-Stokes equations, we obtain
\begin{equation}
\ez (\partial_t - L)\bu_1 + \ez^2 \left[ (\partial_t - L)\bu_2 +  C(\bu_1,\bu_1) + \partial_T \bu_1 \right] + O(\ez^3) = \bm{0},
\label{eq:aaa}
\end{equation}
subject to $\bm{u}_2(0) = \alpha \bu_o$, and $\bm{u}_i(0) = \bm{0}$ for $i\neq 2$. In its primary quality of inverse propagator, the following property holds for $e^{-L t}$: $\partial_t (e^{-Lt}) = -e^{-Lt}L$, where the commutation of $e^{-Lt}$ and $L$ has \textit{not} been used. Thanks to this relation, we write $(\partial_t - L)\bu_i = e^{L t} \partial_t (e^{-L t}\bu_i )$. 
As a result, $L$ disappears from the asymptotic expansion but $e^{-L t}$ appears. The latter is perturbed according to (\ref{eq:pertP}), leading to $e^{L t} \partial_t (e^{-L t}\bu_i ) = e^{L t} \partial_t (\Phi(t) \bu_i ) + \ez e^{L t} \partial_t (P(t) \bu_i)$ for $i=1,2,...$. The asymptotic expansion (\ref{eq:aaa}) becomes
\begin{equation}
\ez e^{L t}\partial_t (\Phi \bu_1)  + \ez^2 \Big[ e^{L t}\partial_t (\Phi \bu_2)  + C(\bu_1,\bu_1) + \partial_T \bu_1 +  e^{L t}\partial_t ( P(t) \bu_1) \Big] + O(\ez^3) = \bm{0}.
\label{eq:bb}
\end{equation}
Note  that the transformation performed from (\ref{eq:aaa}) to (\ref{eq:bb}) is not restricted to time-independent base flows, as the property $\partial_t(\Psi(t)^{-1})= -\Psi(t)^{-1}L(t)$ holds for a time-varying operator $L(t)$ and the associated propagator $\Psi(t)$. This can be shown easily by taking the time derivative of $\Psi(t)^{-1}\bm{u}(t) = \bm{u}(0)$. Terms of (\ref{eq:bb}) are then collected at each order in $\ez$, leading to a succession of linear problems, detailed hereafter.

At order $\ez$, we collect $\partial_t (\Phi \bu_1) = \bm{0}$, subject to $\bu_1(0) = \bm{0}$. We obtain $\Phi \bu_1 = \Phi(0)\bu_1(0) = \bm{0}$, therefore  $\bu_1(t,T)$ is proportional to the kernel of $\Phi(t)$ for all $t \geq 0$. 
We choose the non-trivial solution 
\begin{equation}
\bm{u}_1(t,T) = A(T)H(t)\bl(t), 
\end{equation}
where the initial condition $\bu_1(0)=\bm{0}$ is enforced by  $H(t)$, while the slowly-varying scalar amplitude  $A(T)$ is continuous in  $T$ and modulates the linear trajectory. 
This choice is motivated by the observation that, since $A$ must be constant in time in the linear regime, we expect it to be \textit{weakly} time-dependent in the \textit{weakly} nonlinear regime. We stress that $A(T)$ does not depend \textit{explicitly} on $t$, such that $\partial_t A =0$. Note that the choice $\bm{u}_1(t) = A(t)H(t)\bl(t)$ would also have been possible, and the assumption of the amplitude depending on a slow time scale is made solely to simplify the ensuing calculations.

At order $\ez^2$, we collect
\begin{equation}
\partial_t (\Phi \bu_2)  +  A^2 H e^{-L t}C(\bl,\bl) + H \frac{\mathrm{d} A}{\mathrm{d} T} e^{-L t} \bl  +  A \mathrm{d}_t (H P \bl) = \bm{0},
\label{eq:Sasymtg}
\end{equation}
subject to $\bu_2(0) = \alpha \bu_o$. We used the property $H(t)^2=H(t)$, which will henceforth be understood. The particular solution of (\ref{eq:Sasymtg}) yields
\begin{equation}
\bu_2(t,T)= \bu_2^{(a)}(t) + A(T)^2\bu_2^{(b)}(t) + \frac{\mathrm{d} A(T)}{\mathrm{d} T}\bu_2^{(c)}(t) + A(T)\bu_2^{(d)}(t),
\end{equation}
where
\begin{align*}
&\mathrm{d}_t\left(\Phi \bu_2^{(a)}\right) =\bm{0}, \quad \mathrm{d}_t\left(\Phi \bu_2^{(b)}\right) = - He^{-Lt}C(\bl,\bl), \\
&\mathrm{d}_t \left(\Phi \bu_2^{(c)}\right) = - H e^{-Lt}\bl , \quad \text{and} \quad \mathrm{d}_t\left(\Phi \bu_2^{(d)}\right) = - \mathrm{d}_t\left(H P \bl \right), 
\end{align*}
subject to the initial conditions $\bu_2^{(a)}(0) = \alpha \bu_o$ and $\bu_2^{(b)}(0) = \bu_2^{(c)}(0) = \bu_2^{(d)}(0) = \bm{0}$. 
Time integration can now be performed without ambiguity as all the partial derivatives ($\partial_t ... $) have been replaced by total derivatives ($\mathrm{d}_t...$). 
After time integration between $t=0$ and $t>0$, we obtain a series of problems for $\bu_2^{(a)}$, $\bu_2^{(b)}$,  $\bu_2^{(c)}$ and $\bu_2^{(d)}$:  
\begin{align*}
&\Phi(t) \bu_2^{(a)}(t)=\Phi(0) \bu_2^{(a)}(0) = \alpha \bu_o,
\end{align*}
since $\Phi(0)=I$; 
\begin{align*}
\Phi(t) \bu_2^{(b)}(t)= - \int_{0}^{t} H(s)e^{-Ls}C[\bl(s),\bl(s)] \mathrm{d}s = e^{-Lt}\forcutwo(t), 
\end{align*}
where
\begin{equation}
\frac{\mathrm{d} \forcutwo }{\mathrm{d} t}  = L \forcutwo -  C(\bl,\bl), \quad  
\forcutwo(0) = \bm{0},
\end{equation}
and where we used that the general solution of $\mathrm{d}_t \bm{x} =L\bm{x} + \bm{F}$ is $\bm{x}(t) = e^{Lt}[\bm{x}(0) + \int_{0}^{t}e^{-Ls} \bm{F}(s)\mathrm{d}s]$; 
\begin{align*}
\Phi(t) \bu_2^{(c)}(t) &= - \int_{0}^{t} H(s) e^{-Ls}\bl(s) \mathrm{d}s \\
& = - \int_{0}^{t} H(s) \ez \bu_o \mathrm{d}s= - \ez \bu_o t,
\end{align*}
since $e^{-L t} \bl(t) = \ez \bu_o$ holds by construction; 
and 
\begin{align*}
\Phi(t) \bu_2^{(d)}(t) =  -\left[H(t) P(t) \bl(t) - H(0) P(0) \bl(0)\right] = - \bu_o, 
\end{align*}
since, by construction, $H(t) P(t) \bl(t) = H(t) \bu_o $. Note that the presence of the Heaviside distribution inside the integral is unimportant. Eventually, 
\begin{align}
\Phi \bm{u}_2  &= \alpha \bu_o 
+ A^2 e^{-L t}   \forcutwo
- \ez t \frac{\mathrm{d} A }{\mathrm{d} T} \bu_o -  A \bu_o, \quad t>0.
\label{eq:tgo2}
\end{align}

Invoking again the Fredholm alternative,  (\ref{eq:tgo2}) admits a non-diverging particular solution if and only if its right-hand side is orthogonal to $\ba(t)$ for all $t>0$. This leads to:
\begin{align}
\left \langle \bu_o , \ba(t) \right \rangle \left(\alpha - A \right) + A^2 \left \langle e^{-Lt}\forcutwo(t),\ba(t) \right \rangle - \ez t \frac{\mathrm{d} A}{\mathrm{d} T}\left \langle \bu_o , \ba(t) \right \rangle = 0,  \quad  t>0.
\label{eq:32}
\end{align}
Dividing (\ref{eq:32}) by $\left \langle \bu_o , \ba(t) \right \rangle$ leads to
\begin{equation}
\begin{split}
\left(\alpha - A \right) + \ez A^2  \mu_2(t) - \ez t \frac{\mathrm{d} A}{\mathrm{d} T} = 0, \quad t>0,
\label{eq:322}
\end{split}
\end{equation}
where
\begin{equation}
\mu_2(t) =   \ez^{-1}\frac{\left \langle e^{-Lt} \forcutwo(t) , \ba(t) \right \rangle}{\left \langle \bu_o, \ba(t)  \right \rangle} =  \frac{\left \langle \forcutwo(t) , \bl(t) \right \rangle}{\left \langle \bl(t), \bl(t)  \right \rangle}.
\label{eq:mu2}
\end{equation}
Equation (\ref{eq:322}) is re-expressed as $E(t,T) = 0$ for $t>0$, where $E(t,T) = \left(\alpha - A \right) + \ez A^2 \mu_2(t) - \ez t \mathrm{d}_T A$. Since $\int_{t\rightarrow0}^{t}\partial_s E(s,T)\mathrm{d}s = E(t,T)-E(t\rightarrow0,T)$, solving $E(t,T) = 0$ is equivalent to solving $\partial_t E(t,T)=0$ for $t>0$ subject to $E(t\rightarrow 0,T) = 0$. Thereby, the partial derivative of (\ref{eq:322}) with respect to the short time scale $t$ is taken, leading to 
\begin{equation}
\begin{split}
\ez A^2 \frac{\mathrm{d} \mu_2(t)}{\mathrm{d} t} - \ez \frac{\mathrm{d} A}{\mathrm{d} T} = 0, \quad 0 < t,
\label{eq:34}
\end{split}
\end{equation}
where we have used that $\partial_t A= 0$ by construction since $A=A(T)$ does not \textit{explicitly} depend on $t$. 
Furthermore, the relation (\ref{eq:34}) is subject to $E(t\rightarrow 0,T) = \lim_{t\rightarrow 0} ( \alpha - A ) =0 $ where we have used that $\forcutwo(t \rightarrow 0) = \forcutwo(0) = 0$. 
To be meaningful, equation (\ref{eq:34}) and its initial condition must be re-written solely in terms of $t$, which is done by evaluating $T$ along $T=\ez t$. The \textit{total} derivative of $A$, denoted $d_t A$, is now needed, as it takes into account the \textit{implicit} dependence of $A$ on $t$. By definition, $d_t A = \partial_t A + \ez \partial_T A = \ez d_T A$, such that the final amplitude equation reads
\begin{equation}
\frac{\mathrm{d} A}{\mathrm{d} t} =  \ez A^2\frac{\mathrm{d} \mu_2(t)}{\mathrm{d} t} , \quad \text{with} \quad A(0) = \alpha,
\label{eq:AmpEqtg}
\end{equation}
as $\lim_{t\rightarrow 0} \left( \alpha - A(\ez t) \right) = 0$ implies $ A(t\rightarrow 0) = \alpha$, and the amplitude $A$ is extended by continuity in $t=0$ so as to eventually impose $A(0) = \alpha$. Note that the evaluation in $T=\ez t$ and the passage to the total derivative would lead to indeterminacy in its solution if performed directly in (\ref{eq:322}), since that equation is not subject to any initial condition. Indeed, at linear level for instance, it would yield $\mathrm{d}_t A = (\alpha-A)/t$, which admits the family of solutions  $A(t) = \alpha + C t$, with $C$ an undetermined constant.

We stress that the inverse propagator 
is not needed to solve the amplitude equation (\ref{eq:AmpEqtg}). Just like the original problem considered in this section, (\ref{eq:AmpEqtg}) is unforced and has a non-zero initial condition. 
In the linear regime, $A=\alpha$  for all times, and the linear gain is $\left \| \ez \alpha \bl(t) \right \|/\left \| \alpha \ez^2 \right \| = \left \| \bl(t) \right \|/ \ez$. 
At $t=t_o$, in particular, we recover  that it is equal to $1/\ez$ since $\left \| \bl(t_o) \right \|=\left \| \bv_o \right \|=1$.

In the following, we call Equation (\ref{eq:AmpEqtg}) the  Weakly Nonlinear Nonnormal transient (WNNt) model. It can be corrected with higher-order terms, which requires solving the linear singular system (\ref{eq:tgo2}), as detailed in  Appendix~\ref{appendix:TGFuji}. We show in particular that singular higher-order solutions are orthogonal to the first-order order solution $\bl(t)$. 

\subsection{Application case: the flow past a backward-facing step}
\label{sec:tgBFS}

\begin{figure}
\centering
  \begin{subfigure}[b]{0.65\linewidth}
\includegraphics[trim={1.8cm 0.05cm 1.2cm 0.1cm},clip,width=1\linewidth]{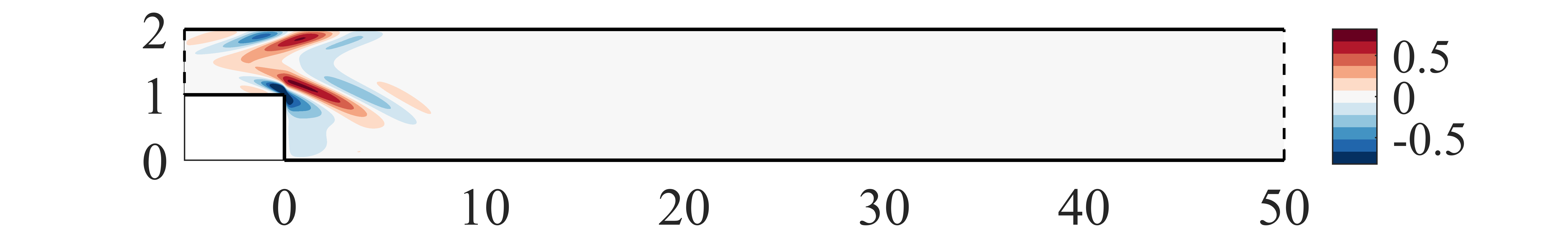}
\caption{\label{fig:fobfs}}
\end{subfigure}
  \hfill
  \begin{subfigure}[b]{0.65\linewidth}
\includegraphics[trim={1.8cm 0.05cm 1.2cm 0.1cm},clip,width=1\linewidth]{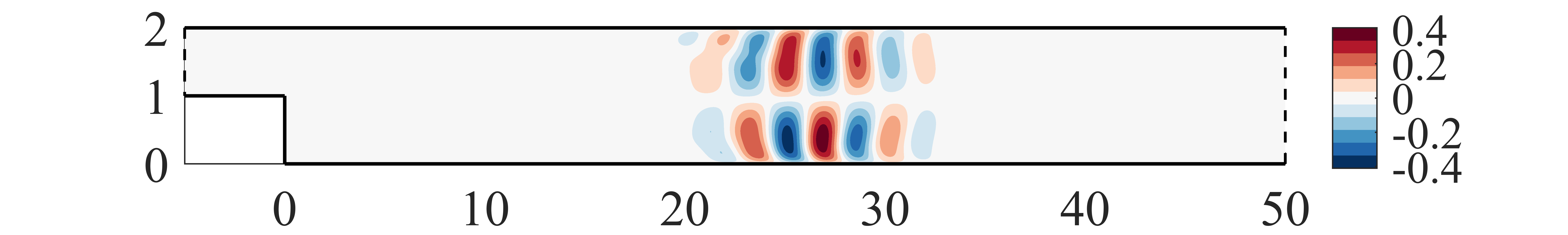}
\caption{\label{fig:uobfs}}
 \end{subfigure}
\caption{\textbf{(a)} Streamwise ($x$) component of the optimal initial condition $\bu_o$ for the BFS (sketched in figure~\ref{fig:sbfs}a) at $Re=500$ and at $t_o = t_{o,m} = 58$
\textbf{(b)}~Streamwise component of the evolution $\bv_o$ at $t=t_o$. Both structures are normalised as $||\bu_o|| = ||\bv_o|| = 1$.\label{fig:bfsoptilin}}
\end{figure}

The WNNt model is applied to the backward-facing step flow for $Re=500$ and $t_o = t_{o,m} = 58$. See  Appendix~\ref{appendix:S3} for details about the numerical method. For these parameters, the linear optimal structures (fig.~\ref{fig:bfsoptilin}) and gain are validated with the results presented in \cite{Blackburn08}. The quadratic term in (\ref{eq:AmpEqtg}), although asymptotically correct, happens to be insufficient to capture the nonlinear saturation of the transient gain for this particular flow, in particular because of the weak value of the coefficient $\mu_2(t)$. Indeed, $\bl(t_o) = \bv_o$ appears to be  dominated by a specific spatial wavenumber (see figure~\ref{fig:uobfs}), thus the field $\forcutwo$, being generated by the nonlinear interaction of $\bl(t)$ with itself, it is dominated by spatial harmonics and its projection on $\bl(t)$ is close to zero. For this flow the WNNt model therefore needs to be extended to order $\ez^3$ (see Appendix~\ref{appendix:TGFuji}), yielding
\begin{equation}
 \frac{\mathrm{d} A}{\mathrm{d} t} = \ez A^2 \frac{\mathrm{d} \mu_2 }{\mathrm{d} t} + \ez^2 A^3 \frac{\mathrm{d} \mu_3 }{\mathrm{d} t}, \quad A(0)=\alpha,
 \label{eq:tgfuji}
\end{equation}
where
\begin{equation}
\mu_3(t) \doteq \frac{\left \langle \forcuthree(t) , \bl(t) \right \rangle}{\left \langle \bl(t), \bl(t)  \right \rangle},
\label{eq:coeffo3tg}
\end{equation}
and
\begin{equation*}
\frac{\mathrm{d} \forcuthree}{\mathrm{d} t}  = L\forcuthree - 2\left[C(\bl,\forcutwo) - \mu_2 C(\bl,\bl) + \dot{\mu}_2(\forcutwo- \mu_2\bl) \right], \quad  \forcuthree(0) = \bm{0}. 
\end{equation*}
Equation (\ref{eq:tgfuji}) is similar to (\ref{eq:AmpEqtg}), although \textit{corrected} by a cubic term. We formulate the amplitude equation (\ref{eq:tgfuji}) in terms of the rescaled quantities $a = \ez A$ and the amplitude of the initial condition $U_0 =||\bm{U}(0)||=\alpha \ez^2$:
\begin{equation}
\frac{\mathrm{d} a}{\mathrm{d} t} = a^2\frac{\mathrm{d}\mu_2}{\mathrm{d} t} + a^3\frac{\mathrm{d}\mu_3}{\mathrm{d} t},\quad a(0)=\frac{U_0}{\ez}.
\label{eq:tgfujiaa}
\end{equation}
\begin{figure*}
\centering
  \begin{subfigure}[b]{0.49\linewidth}
\includegraphics[trim={3.cm 9.5cm 4cm 9.5cm},clip,width=1\linewidth]{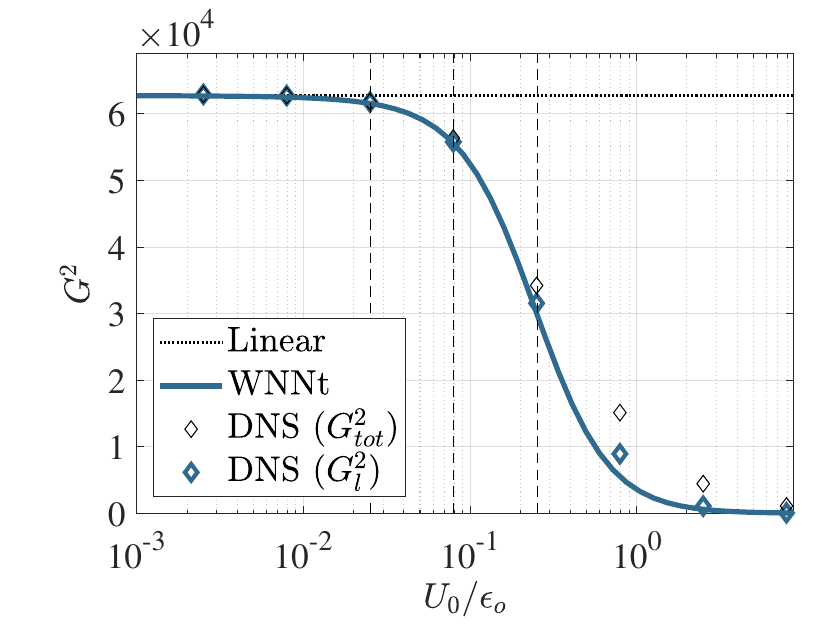}
\caption{\label{fig:bfstg_a}}
\end{subfigure}
  \hfill
    \begin{subfigure}[b]{0.49\linewidth}
\includegraphics[trim={3.0cm 9.5cm 4cm 9.5cm},clip,width=1\linewidth]{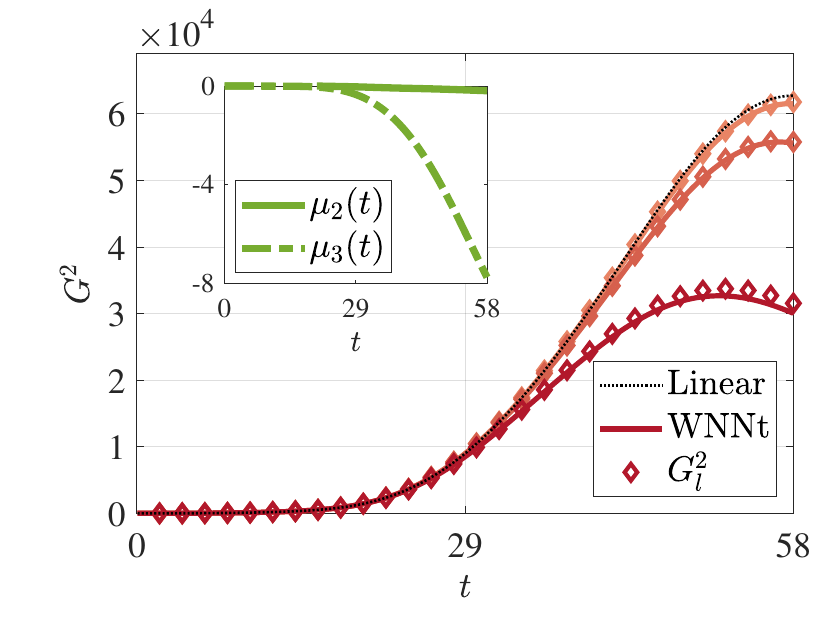}
\caption{\label{fig:bfstg_b}}
\end{subfigure}
\caption{
Transient gain in the flow past a backward-facing step (sketched in figure~\ref{fig:spois}a) for $Re = 500$.
\textbf{(a)} Gain squared $G(t_o)^2$ for $t_o=t_{o,m}=58$ as a function of the amplitude of the initial condition. 
\textbf{(b)}~History of the gain squared for $0\leq t \leq t_{o,m}$ and for three amplitudes of initial condition, $U_0/\ez = [0.025,0.08,0.25]$ (vertical dashed lines in (a)); larger amplitudes darker. Inset: weakly nonlinear coefficients $\mu_2(t)$ (continuous line) and $\mu_3(t)$ (dashed-doted line) as a function of time. 
\label{fig:WNLtgBFS}}
\end{figure*}
In this manner, the weakly nonlinear transient gain becomes $G(t_o) = a(t_o)/U_0$. Note that in (\ref{eq:tgfujiaa}) the amplitude $a(t)$ does not depend on $U_0$ nor on $\ez$ independently, but on their ratio $U_0/\ez$. Thus, as expected, increasingly nonlinear regimes are found when the amplitude of the initial condition increases with respect to the linear gain.

Predictions from equation (\ref{eq:tgfujiaa}) are shown in figure~\ref{fig:WNLtgBFS} together with the linear and fully nonlinear DNS gains evaluated in two ways: using either the total perturbation around the base flow, for $G_{tot}$ with
\begin{equation*}
G_{tot}(t) = \frac{|| \bm{U}(t) - \bm{U}_e ||}{U_0}, 
\end{equation*}
or using that perturbation projected on $\bl(t)$, for $G_{\bl}$. Namely, we recall the asymptotic expansion of the solution $\bm{U} = \bm{U}_e + a \bl + \ez^2 \bu_2 + O(\ez^3)$  for $t>0$, with $\left \langle  \bl,\bu_i \right \rangle = 0$ for all $i \geq 2$ as a consequence of the Fredholm alternative (see Appendix \ref{appendix:TGFuji}). This gives: 
\begin{equation*}
a(t) = \frac{\left \langle \bm{U} - \bm{U}_e , \bl(t) \right \rangle}{||\bl(t)||^2}. 
\end{equation*}
Thus, from the  knowledge of $\bm{U}-\bm{U}_e$ computed by DNS, the gain that should be compared with the weakly nonlinear prediction $a ||\bl(t)||/U_0 $ is
\begin{equation*}
G_{\bl}(t) = \frac{\left \langle \bm{U}(t) - \bm{U}_e , \bl(t) \right \rangle}{U_0 ||\bl(t)||}, 
\end{equation*}
which, evaluated at $t=t_o$ (with $\bl(t_o) = \bv_o$ of unit norm), gives $G_{\bl}(t_o)=\left \langle \bm{U}(t_o) - \bm{U}_e , \bv_o \right \rangle/U_0$.
In figure~\ref{fig:bfstg_a}, the WNNt model extended to $\ord(\ez^3)$ appears to capture the weakly evolution of the transient gain with precision, in particular $G_{\bl}$. When $G_{\bl}$ and $G_{tot}$ depart from each other, the higher-order fields $\forcutwo$, $\forcuthree$, $\ldots$ are expected to have a significant amplitude, and thus the WNNt prediction deteriorates since it is based on an asymptotic hierarchy. For this specific flow, however, the error remains small and the prediction  satisfactory even in the fully nonlinear regime. In figure~\ref{fig:bfstg_b}, the gain history of $G_{\bl}$ for all times $0 \leq t \leq t_o$ is successfully compared to $a(t)||\bl(t)||/U_0$. The coefficient $\mu_3(t)$ is much larger than $\mu_2(t)$ (inset), and is largely dominated by the part of $\forcuthree$ generated by the forcing term $C(\bl,\forcutwo)$. Since $\mu_3(t)$ is monotonously decreasing toward $\mu_3(t_o) = -7.77$, larger times are subject to a stronger saturation. This leads to a decrease of the time for which the specific initial condition $\bu_o$ leads to a maximum transient gain, consistently with the DNS results.    

\subsection{Application case: Lift-up in the plane Poiseuille flow}
\label{sec:tgPoiseuille}

\begin{figure}
\centering
   \begin{subfigure}[b]{0.49\linewidth}
\includegraphics[trim={0.cm 2.5cm 0cm 3.5cm},clip,width=1\linewidth]{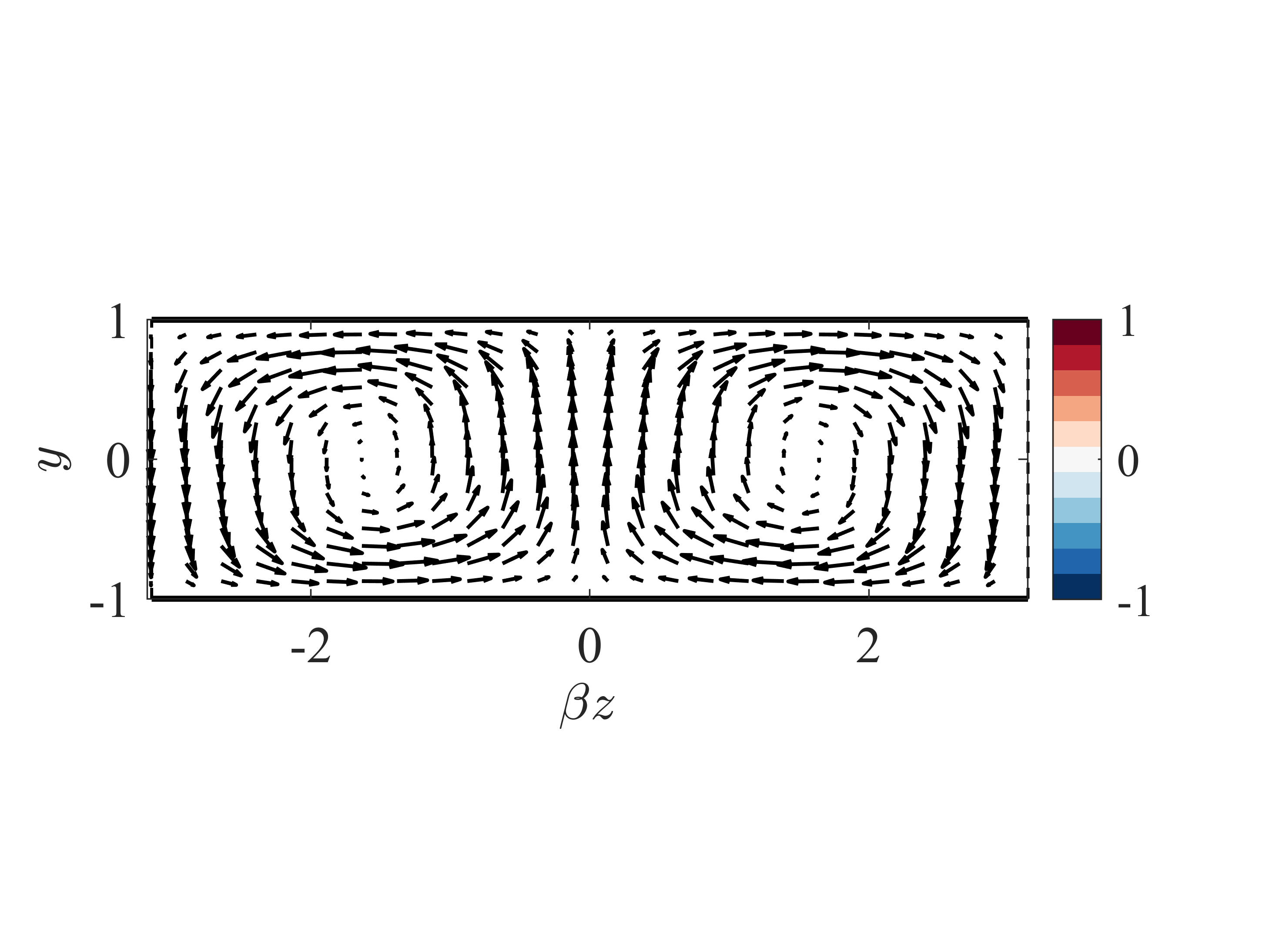}
\caption{\label{fig:fopois}}
\end{subfigure}
  \hfill
    \begin{subfigure}[b]{0.49\linewidth}
\includegraphics[trim={0.0cm 2.5cm 0cm 3.5cm},clip,width=1\linewidth]{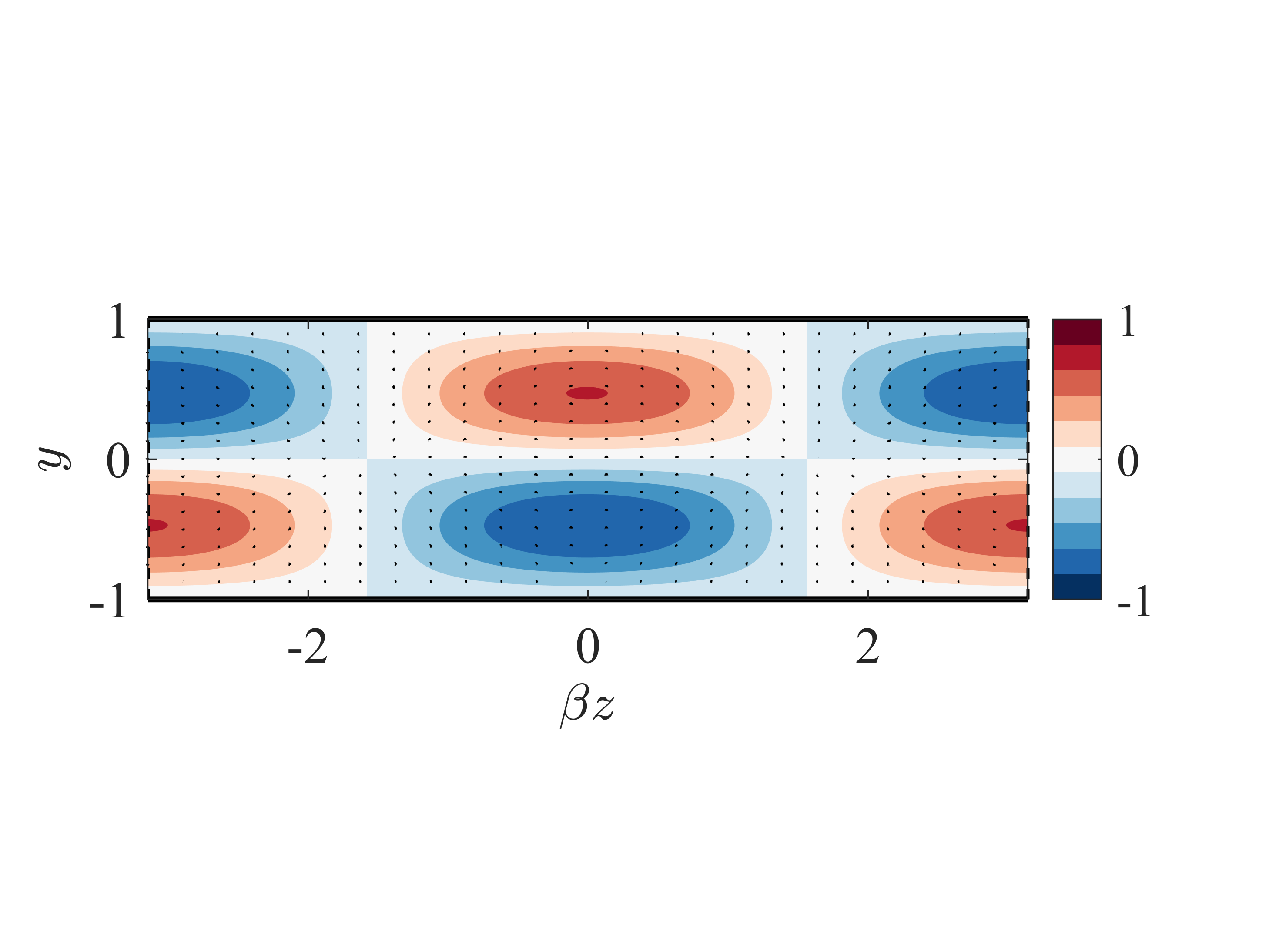}
\caption{\label{fig:uopois}}
\end{subfigure}
\caption{\textbf{(a)} Optimal initial condition $\bu_o$ for the plane Poiseuille flow (sketched in figure~\ref{fig:spois}b) for $(Re,k_x,k_z) = (3000,0,2)$ and $t_o=t_{o,m}=230$. 
Arrows: cross-sectional velocity field ($u_{o,z}$,$u_{o,y}$). 
Contours: streamwise  component $u_{o,x}$.
\textbf{(b)}~Evolution $\bv_o$ at $t=t_o$. Both fields are normalised as  $||\bu_o|| = || \bv_o|| = 1$. 
Initial vortices have a null streamwise component $u_{o,x}$, and streaks at $t=t_o$ have negligible cross-sectional components ($v_{o,z}$,$v_{o,y}$). Only one wavelength  $-\pi \leq \beta z \leq \pi$ is shown.}
\end{figure}

The WNNt is now applied to the plane Poiseuille flow. The set of parameters $(Re,k_x,k_z,t_o) = (3000,0,2,t_{o,m}=230)$ is selected. In both the linear and nonlinear computations, the wavenumber $k_x=0$ is maintained such that the fields are constant in $x$, and only the dependence in $y$ and $z$ is computed. Contrarily to the application case \S\ref{sec:harmonicPoiseuille}, perturbations can now be fully three dimensional (i.e. $\bu=(u_x(y,z),u_y(y,z),u_z(y,z))$. The computations are performed in the spanwise-periodic box $(y,z) \in [-1,1] \times [-\pi/k_z,\pi/k_z] \equiv  \Omega$. All the scalar products are taken upon integration inside this periodic box, in particular for the normalisation $\langle \bu_o,\bu_o\rangle = \langle \bv_o,\bv_o\rangle =1$, and for the evaluation of the weakly nonlinear coefficients. The linear optimal gain is validated with the result of \cite{SH01}; the associated optimal initial condition and its evolution at $t=t_o$ are shown in figures~\ref{fig:fopois} and \ref{fig:uopois}, respectively. The optimal initial condition consists of vortices aligned in the streamwise direction; as these streamwise vortices are superimposed on the parabolic base flow, they bring low-velocity fluid from the wall towards the channel centre and high-velocity fluid from the centre of the channel towards the walls, thus generating alternated streamwise streaks.
Due to the spanwise periodicity of the optimal initial condition $\bu_o$, all the  solutions at \textit{even} orders $\ez^{2n}$ $(n=1,2,3,...)$  only yield \textit{even} spatial harmonics in $k_z$
and are orthogonal to $\bl(t)$, such that the coefficient $\mu_2(t)$ defined in (\ref{eq:mu2}) is null at all times. Therefore, (\ref{eq:tgfujiaa}) reduces to
\begin{equation}
\frac{\mathrm{d} a}{\mathrm{d} t} =  a^3\frac{\mathrm{d} \mu_3 }{\mathrm{d} t},\quad a(0)=\frac{U_0}{\ez},
\label{eq:AmpEqtg_o3}
\end{equation}
and $\forcuthree$ solves the simplified equation
\begin{equation}
\frac{\mathrm{d} \forcuthree }{\mathrm{d} t}  = L\forcuthree -  2C(\forcutwo,\bl), \quad  
\forcuthree(0) = \bm{0}.
\end{equation}
The analytical solution of (\ref{eq:AmpEqtg_o3}) writes
\begin{equation}
a(t)= \frac{U_0}{\ez}\left[ 1-\left( \frac{U_0}{\ez} \right)^2 2\mu_3(t) \right]^{-1/2}.
\label{eq:AmpEqtgo3}
\end{equation}
We show in Appendix~\ref{appendix:S11} that, at first order in the gain variation,  (\ref{eq:AmpEqtgo3}) reduces to the sensitivity of the transient gain to the  base flow modification $(U_0/\ez)^2\forcutwo(t)$.

\begin{figure*}
\centering
  \begin{subfigure}[b]{0.49\linewidth}
\includegraphics[trim={3.cm 9.5cm 4cm 9.5cm},clip,width=1\linewidth]{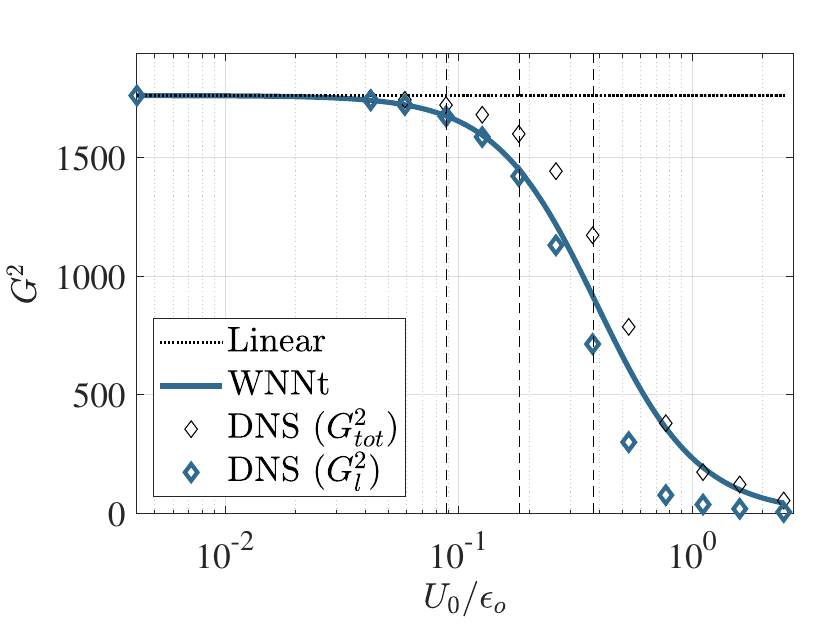}
\caption{\label{fig:pois_a}}
\end{subfigure}
  \hfill
    \begin{subfigure}[b]{0.49\linewidth}
\includegraphics[trim={3.0cm 9.5cm 4cm 9.5cm},clip,width=1\linewidth]{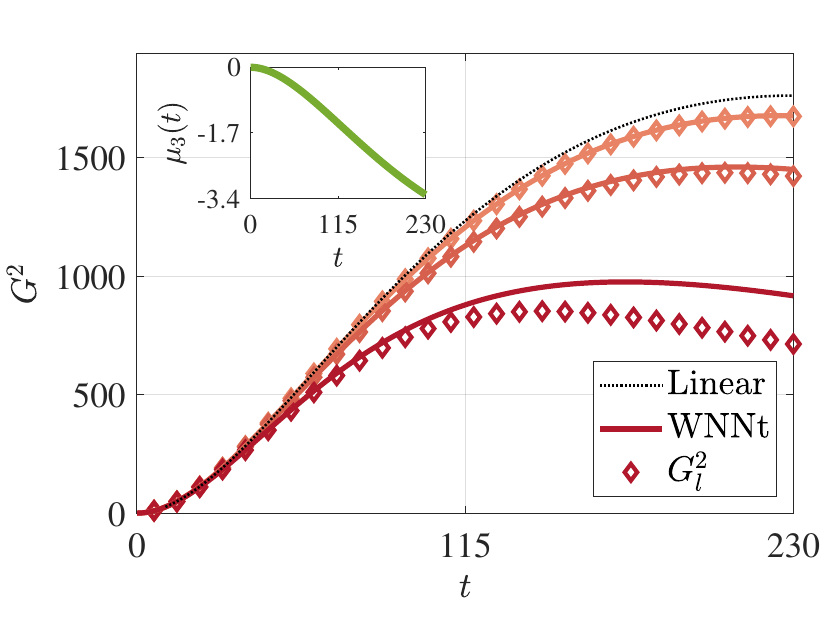}
\caption{\label{fig:pois_b}}
\end{subfigure}
\caption{
Transient gain in the plane Poiseuille flow (sketched in figure~\ref{fig:spois}b) for $(Re,k_x,k_z) = (3000,0,2)$.
\textbf{(a)} Gain squared $G(t_o)^2$ for $t_o=t_{o,m}=230$ as a function of the amplitude of the initial condition. 
Streamwise invariance $k_x=0$ is enforced in the DNS as well. 
\textbf{(b)}~History of the gain squared for $0\leq t \leq t_{o,m}$ and for three amplitudes of initial condition, $U_0/\ez = [0.088,0.18,0.37]$ (vertical dashed lines in (a)); larger amplitudes darker. Inset: weakly nonlinear coefficient $\mu_3(t)$ as a function of time. 
\label{fig:WNLtgPois}}
\end{figure*}

Predictions from equation (\ref{eq:AmpEqtgo3}) are shown in figure~\ref{fig:WNLtgPois} together with the linear and fully nonlinear DNS gains.  
The WNNt model predicts  $G_{\bl}$  accurately in the weakly nonlinear regime for $t=t_{o,m}$, which supports our approach (figure~\ref{fig:pois_a}). In the  strongly nonlinear regime, beyond $U_0/\ez \approx 0.4$, the model overestimates $G_{\bl}$. Again, this can be interpreted by noting that
$G_{tot}$  is twice as large as $G_{\bl}$, i.e. more energy is contained in the higher-order terms generated by the linear response than in the linear response itself. 
Therefore, this is not the amplitude equation (\ref{eq:AmpEqtgo3}) that breaks down, but the very idea of an asymptotic expansion. Whether higher-order terms remain smaller than the fundamental is certainly flow-dependent, and the WNNt model is expected to be even more accurate when this is the case, as shown in \S\ref{sec:tgBFS} for the flow past a backward-facing step, which generated rather weak higher-order fields.

Figure \ref{fig:pois_b} compares  for $t \leq t_o$ the history of the approximated gain $a(t) ||\bl(t)||/U_0$ with that of the DNS gain $G_{\bl}$, and shows a convincing overall agreement. The coefficient $\mu_3(t)$ is negative and decays monotonously with time until $\mu_3(t_o) = -3.30$ (inset), enhancing the saturation. This results in a reduction of the approximated optimal time with forcing amplitude and associated with the initial condition $\bu_o$, as also observed in the DNS.

\section{Conclusions}

In summary, we have derived two weakly nonlinear  amplitude equations for nonnormal systems,  describing the asymptotic response to harmonic forcing and the transient response to initial condition.
In both cases, the presence of a neutral or weakly damped mode is unnecessary. 
Both approaches are based on the same observation: in nonnormal systems, the resolvent and  propagator operators may lead to a notable amplification, so their inverses may in contrast lead to a notable \textit{mitigation} of the response and are close to singular. 
A small perturbation is then sufficient to fill their kernel with the response and render them singular. This can be encompassed in a multiple-scale expansion, closed by means of classical compatibility conditions. 

The resulting amplitude equations have been validated with fully nonlinear simulations, both in parallel and non-parallel two-dimensional flows. 
In all cases, they predict accurately the supercritical or subcritical nonlinear evolution of the response, and bring insight on the weakly nonlinear mechanisms that modify the gains as the amplitude of the harmonic forcing or the initial condition varies. 
In particular, the efficiency of the WNNh model to capture a subcritical behaviour may prove useful in the search of optimal paths to chaos or turbulence. 
Indeed, equation (\ref{eq:ea}) could be included as an additional constraint in a Lagrangian optimisation problem, whose stationary point would constitute a \textit{weakly-nonlinear optimal}. Such an approach could complement fully nonlinear optimisations (\cite{Pringle10}), by providing  physical understanding at a numerical cost close to the linear one.

It should be noted that the proposed method is not restricted to the Navier-Stokes equations, but apply to all nonlinear systems whose linearised operator exhibits a strong nonnormality (see \cite{Tref05} §55-60 for a comprehensive discussion, as well as the situations discussed in the introduction). For instance in ecological models describing the temporal evolution of a population, such as the canonical Lotka-Volterra predator-prey equations, the so-called \textit{resilience} of a community (spectral abscissa of the Jacobian of the system) is known to be sometimes a misleading or incomplete measure \cite{Neubert97}. The conjunction of  nonnormality and nonlinearity is then key to predict the extinction of a population.

The amplitude equations proposed in the present study are expected to be relevant for three-dimensional flows. 
For the transient growth aspect, the assumption of a time-independent state operator and the associated operator exponential formalism are unnecessary, and we believe that the model can be extended to time-varying base flows.

Another interesting extension of the models, apart from higher-order corrections, is the inclusion of multiple forcing structures or trajectories, originating for instance from additional singular vectors in the asymptotic expansions.
The nonlinear interaction of multiple harmonic forcings or initial conditions is particularly relevant when distinct structures lead to comparable gains, for instance optimal and sub-optimal initial conditions \cite{Butler92, Blackburn08}, or perturbations of different spatial wavenumbers like in  jet flows forced with different azimuthal wavenumbers \cite{Garnaud13B}.
The ensuing system of coupled amplitude equations may  bear rich dynamics, such as hysteresis and chaos. 
Our current efforts also involve deriving an amplitude equation for the response to stochastic forcing, as investigated in \cite{Farrell93} and \cite{Lugo16s} with linear and self-consistent models, respectively. 


\appendix

\section{Harmonic gain sensitivity and comparison with the WNNh model. \label{appendix:S2}}
Let $G_o = 1/\ez$ designate the linear harmonic gain. Then $R^\da R \hbf_o = G_o^2 \hbf_o$ holds by definition, and implies $G_o^2 = \left \langle  R^\da R \hbf_o, \hbf_o \right \rangle$ thanks to the chosen normalisation $||\hbf_o||=1$. We are interested in the 
squared gain variation $\de G_o^2$ (where this notation does \textit{not} designate the square of the gain variation) induced by a small perturbation $\de L$ of the state operator. The latter results in the following perturbation $\de R$ of the resolvent:
\begin{align*}
\de R &= (i\w I-L-(\de L))^{-1} - ( i\w I-L)^{-1} \\
&=\left[ (i\w I-L)(I - R (\de L)) \right]^{-1}  - R \\
& \approx (I+R\de L)R-R \\
&= R (\de L) R.
\end{align*}
The gain variation is therefore
\begin{align*}
\de G_o^2 &= \langle \de(R^\da R)\hbf_o,\hbf_o\rangle = \left \langle (\de R)^\da R \hbf_o + R^\da(\de R)\hbf_o ,  \hbf_o \right \rangle 
\\
&= \left \langle R^\da(\de R)\hbf_o  ,  \hbf_o \right \rangle + c.c 
\\
&= \left \langle R (\de L) R\hbf_o  ,  R \hbf_o \right \rangle + c.c \\
&= \left \langle (\de L) G_o \hbu_o , G_o^2 \hbf_o  \right \rangle + c.c.,
\end{align*}
so finally
\begin{align}
\de G_o^2 = 2G_o^3 \Re\left[\left \langle (\de L)  \hbu_o ,\hbf_o  \right \rangle\right].
\label{eq:perp}
\end{align}
For instance, a base flow modification $\de \bm{U}_e$ results in $(\de L)\hbu_o = -(\hbu_o \cdot \nabla) \de \bm{U}_e - (\de \bm{U}_e \cdot \nabla)\hbu_o = -2 C(\hbu_o,\de \bm{U}_e)$ yielding the same formula as in \cite{Brandt11} with a different normalisation.

On the other hand, the WNNh model predicts: 
\begin{align}
 Y^3\frac{ \left | \mu + \nu \right |^2 }{\ez^2 \left | \eta \right |^2} + 2 Y^2 \Re\left[\frac{\mu + \nu}{\ez\eta}\right] + Y  = \left( \frac{F}{\ez} \right)^2. 
 \label{eq:S_Sea}
\end{align}
where $Y=|\bar{a}|^2$. We identify the weakly nonlinear harmonic gain as $G^2 =Y/F^2$, and multiply (\ref{eq:S_Sea}) by $\ez^2/Y$:
\begin{align}
 Y^2\frac{ \left | \mu+\nu \right |^2 }{\left | \eta \right |^2} + \frac{2 Y}{G_o} \Re\left[\frac{\mu + \nu}{\eta}\right] + \frac{1}{G_o^2}  - \frac{1}{G^2} = 0.
\end{align}
Being interested in small variations around $G_o^2$ (that correspond to the linear limit $Y = |\bar{a}|^2 \rightarrow 0$), we write $G^2 = G_o^2 + \de G_o^2$ with $|\de G_o^2/G_o^2| \ll 1$. In this manner, $1/G_o^2-1/G^2 = \de G_o^2 /G_o^4 + h.o.t$, eventually leading to 
\begin{align}
\de G_o^2 = -2G_o^3\Re\left[\frac{|\bar{a}|^2(\mu+\nu)}{\eta}\right] + h.o.t  .
\end{align}
We recognise at leading order equation (\ref{eq:perp}) where $(\de L)\hbu_o = - |\bar{a}|^2 [ 2C\left( \hbu_o,\bu_{2,0} \right) + 2C\left( \hbu_o^*,\hbu_{2,2} \right) ] $. Thus, in the small gain variation limit, the WNNh model both contains the sensitivity formula of the harmonic gain to the base flow static perturbation $|a|^2 \bu_{2,0}$, and embeds the effect of the second harmonic $\hbu_{2,2}$ as well.
\section{Applying the WNN models to the Navier--Stokes equations. \label{appendix:S3}}

The incompressible Navier--Stokes  equations  write after linearising around the equilibrium velocity field $\bm{U}_e$
\begin{equation}
\begin{split}
B\frac{\mathrm{d} \tq}{\mathrm{d} t} = L \tq + \td,
\end{split}
\end{equation}
with the state vector $\tq=[\bm{u},p]^T$, the forcing $\td=[\bm{f},0]^T$, the singular mass matrix
\begin{equation*}
B = \begin{bmatrix}
I & 0\\ 
0 & 0
\end{bmatrix}, \quad
\end{equation*} 
and the linearised Navier--Stokes operator
\begin{equation*}
L = \begin{bmatrix}
- (\bm{U}_e \cdot \nabla) * - (* \cdot \nabla) \bm{U}_e  + Re^{-1}\Delta (*)   \ & \nabla (*) \\ 
\nabla \cdot (*) & 0
\end{bmatrix}. 
\label{eq:defls}
\end{equation*} 
Several subtleties arise from the peculiarity of the pressure variable, that ensures the instantaneous satisfaction of the incompressibility condition: (i) the absence of time-derivative of the pressure results in a singular mass matrix, 
(ii) forcing terms remain restricted to the momentum equations as we choose to have no source/sink of mass
and (iii) the pressure is not included in the energy norm of the response. This complicates slightly the practical computation of the gain. For the harmonic response model, the resolvent operator is generalised as $R(i\wz) = \left(i \wz B - L\right)^{-1}$, and the gain is measured according to 
\begin{equation*}
G^2(i \wz) = \frac{ \langle \htq , \htq \rangle_B }{\langle \htd , \htd \rangle},
\end{equation*}
where we used the following scalar products
\begin{equation*}
\begin{split}
\langle \hbq_a, \hbq_b \rangle_B &= \int_{\Omega}^{} \left( \hu_{a,x}^*\hu_{b,x} + \hu_{a,y}^*\hu_{b,y} + \hu_{a,z}^*\hu_{b,z} \right) \mathrm{d}\Omega, \quad \text{and} \\
\langle \hbq_a, \hbq_b \rangle  &= \int_{\Omega}^{}\left( \hu_{a,x}^*\hu_{b,x} + \hu_{a,y}^*\hu_{b,y} + \hu_{a,z}^*\hu_{b,z} + \hat{p}_{a}^*\hat{p}_{b} \right)\mathrm{d}\Omega. \\
\end{split}
\end{equation*} 
The $B$-scalar product excludes  pressure, such that the pseudonorm $\langle \hbq , \hbq \rangle_B = ||\hbq||_B^2$ is the total \textit{kinetic} energy of the response. The scalar product at the denominator includes pressure, although this will not change the norm of $\hbd$, namely $\langle \hbd, \hbd \rangle = ||\hbd||^2$, since we have no source/sink of mass. The weakly nonlinear coefficients must be considered under these scalar products. Let $\hbq_o = [\hbu_o,p_o]^T$ with $||\hbq_o||_B=1$, $\hbd_o = [\hbf_o,0]^T$ with $||\hbd_o||=1$, and $\hbd_h = [\hbf_h,0]^T$ with $||\hbd_h||=1$, then: 
\begin{equation*}
\begin{split}
& \gamma = \langle  \hbd_o , \hbd_h  \rangle, \quad 1/\eta = \langle  \hbd_o ,  B \hbq_o \rangle  =   \langle  \hbd_o ,  \hbq_o \rangle_B, \\
&\mu/\eta = \langle  \hbd_o  , \hbd_{3,1}^{(3)} \rangle , \quad 
\nu/\eta =  \langle  \hbd_o  ,\hbd_{3,1}^{(4)} \rangle,
\end{split}
\end{equation*}
where 
$\hbd_{3,1}^{(3)} = [\hbf_{3,1}^{(3)},0]^T$,
$\hbd_{3,1}^{(4)} = [\hbf_{3,1}^{(4)},0]^T$, 
$\hbf_{3,1}^{(3)} = 2C(\hbu_o,\bu_{2,0})$ and 
$\hbf_{3,1}^{(4)} = 2C(\hbu^*_o,\hbu_{2,2})$. 
The pressure field has no influence on the weakly nonlinear coefficients. For instance
\begin{equation*}
\begin{split}
1/\eta &=  \int_{\Omega}^{} \left( \hf_{o,x}^*\hu_{o,x}+ \hf_{o,y}^*\hu_{o,y}+ \hf_{o,z}^*\hu_{o,z} \right) \mathrm{d}\Omega,  \quad \text{and} \\
\mu/\eta  &= \int_{\Omega}^{} \left( \hf_{o,x}^*\hf_{(3,1),x}^{(3)}+ \hf_{o,y}^*\hf_{(3,1),y}^{(3)}+ \hf_{o,z}^*\hf_{(3,1),z}^{(3)} \right) \mathrm{d}\Omega.
\end{split}
\end{equation*}

The linear and nonlinear Navier-Stokes equations are solved for ($u_x$,$u_y$,$p$) by means of the Finite Element Method with Taylor-Hood (P2, P2, P1) elements, respectively, after implementation of the weak form in the software FreeFem++. The steady solutions of the Navier-Stokes equations are solved using the iterative Newton–Raphson method, and the linear operators are built thanks to a sparse solver implemented in FreeFem++. The singular value decomposition is performed in Matlab following directly \cite{Garnaud13B}. Finally, DNS are performed by applying a time scheme based on the characteristic–Galerkin method as described in \cite{Benitez11}. \newline
For the two-dimensional flow past a BFS presented in \S\ref{sec:harmonicBFS}, we refer to \cite{Lugo16} for the validation of the codes with existing literature and the mesh convergence,  since the same codes have been used. 
The length of the outlet channel is chosen as $L_{out}=50$ for $Re \leq 500$ \citep{Lugo16},  $L_{out}=65$  for $Re=600$, and $L_{out} = 80$ for $Re=700$. This ensures the convergence of the linear gain and weakly nonlinear coefficients. For the plane Poiseuille studied in \S\ref{sec:harmonicPoiseuille}, the validation is proposed in the main text. \\

For the transient growth model, the linearised problem writes
\begin{equation}
B\frac{\mathrm{d} \tq}{\mathrm{d} t} = L \tq \quad \text{subject to} \quad  \tq(0) = \begin{bmatrix}
\bu(0)\\ 
0
\end{bmatrix},
\label{eq:tgd}
\end{equation}
and the gain is measured as
\begin{equation}
G(t_o)^2 = \frac{ \langle \tq(t_o), \tq(t_o) \rangle_B}{ \langle \tq(0), \tq(0) \rangle_B} ,
\label{eq:gtgmass}
\end{equation}
where the pressure component of the initial condition can be chosen as $p(0)=0$. The orthogonality properties holds under the $B$-scalar product, and the weakly nonlinear coefficient $\mu_2(t)$ writes 
\begin{equation}
\mu_2(t)  =  \ez \frac{\left \langle \tilde{\bm{q}}_2(t) , \tq_{\bl}(t) \right \rangle_B}{\left \langle  \tq_{\bl}(t),  \tq_{\bl}(t)  \right \rangle_B},
\end{equation}
where $ \tq_{\bl}(t) = [\bl(t),p_{\bl}(t)]^T$,
and where $\tilde{\bm{q}}_2  = [\forcutwo(t),\tilde{p}_2(t)]^T $ is solution of
\begin{equation}
B \frac{\mathrm{d} \tilde{\bm{q}}_2}{\mathrm{d} t}  = L\tilde{\bm{q}}_2 - \begin{bmatrix}
C(\bl,\bl)\\ 
0
\end{bmatrix}, \quad  \tilde{\bm{q}}_2(0) = \bm{0}.
\end{equation}
Again,  pressure does not influence the weakly nonlinear coefficient since only  velocity fields are involved in the scalar product. In particular at $t=t_o$, $\tq_{\bl}(t_o)= [\bv_o,p_{\bl}(t_o)]^T$ 
thus $\left \langle \tq_{\bl}(t_o),\tq_{\bl}(t_o)\right \rangle_B =1$ by construction, and
\begin{equation}
\mu_2(t_o)  =  \ez \int_{\Omega}^{} \tilde{u}_{2,x} v_{o,x} + \tilde{u}_{2,y} v_{o,y} + \tilde{u}_{2,z} v_{o,z} \mathrm{d}\Omega.
\end{equation}
The software FreeFem++ is again used to solve for the velocity and pressure by means of the Finite Element Method with Taylor-Hood elements, (P2 for  velocity and P1 for  pressure). The practical computation of the gain (\ref{eq:gtgmass}) proposed in \cite{Garnaud13} is followed. 
The application of the propagator $e^{Lt}$ (resp. its adjoint $(e^{Lt})^\da$) are performed by integrating in time the linearised problem (\ref{eq:tgd}) (resp. the adjoint problem) with the Crank-Nicolson method. The application of the inverse propagator $e^{-Lt}$ is never needed. \newline
For the transient growth past the backward-facing step studied in \S\ref{sec:tgBFS}, our linear optimisation codes are validated upon comparison with the results of \cite{Blackburn08}. For $(Re,t_o) = (500,58)$, we obtained $G(t_o)^2 = 62.8 \times 10^{3}$, against $G(t_o)^2 = 63.1\times 10^{3}$ in \cite{Blackburn08}. The $\approx 0.5 \%$ relative error could be explained by the fact that our entrance length is $L_i = 5$, against $L_i=10$ in \cite{Blackburn08}. For the plane Poiseuille flow analysed in \S\ref{sec:tgPoiseuille}, the validation was performed thanks to the open-source results of \cite{SH01}, obtained with a Chebyshev polynomial discretisation, and where the singular value decomposition of the matrix exponential $e^{Lt}$ is performed directly. For the chosen set of parameters $(Re,k_x,k_z,t_o) = (3000,0,2,230)$, convergence was achieved for a squared linear gain of $G(t_o)^2 = 1761.8$, against $G(t_o)^2 = 1761.9$ in \cite{SH01}.
\section{\label{appendix:Pois1} Higher-order corrections of the WNNh equation.}

Recall the equation (\ref{eq:ro3}) obtained at order $\sqrt{\ez}^3$:
\begin{align}
\Phi \obu_{3,1} = - A|A|^2 \left[ 2C(\hbu_o,\bm{u}_{2,0}) + 2C(\hbu_o^*,\hbu_{2,2}) \right] - \hbu_o\frac{\mathrm{d} A}{\mathrm{d} T} - A \hbf_o + \phi \hbf_h, 
\label{eq:o3ls_bf}
\end{align}
After imposition of the Fredholm alternative, leading to the equation (\ref{eq:wnlr}) for $\mathrm{d} A/\mathrm{d}T$, the relation (\ref{eq:o3ls_bf}) becomes:
\begin{equation}
\begin{split}
\Phi \obu_{3,1} =& A|A|^2 \left[ -2C(\hbu_o,\bm{u}_{2,0}) - 2C(\hbu_o^*,\hbu_{2,2}) + \zeta \hbu_o \right] \\ 
& + A(-\hbf_o + \eta\hbu_o ) + \phi(\hbf_h - \eta \gamma\hbu_o )
\end{split}
\label{eq:o3ls_af}
\end{equation}
where $\zeta = (\mu+\nu)$. For higher-order corrections of the WNNh model, the field $\obu_{3,1}$ is needed and is solution of (\ref{eq:o3ls_af}) where the operator $\Phi$ is singular since $\hbu_o \neq \bm{0}$ belongs to its kernel. Since by construction $\hbf_o = \ez R(i\wz)^\da \hbu_o$, it follows immediately that $\langle \mbox{RHS} ,\hbf_o \rangle = \ez \langle R(i\wz) \mbox{RHS} ,\hbu_o \rangle$. 
Thus, thanks to the (imposed) orthogonality of the RHS with $\hbf_o$ ($\langle \mbox{RHS} ,\hbf_o \rangle=0$), solving the equation replacing $\Phi $ by $(i\wz I-L)$ leads directly to $\obu_{3,1}$ being orthogonal to $\hbu_o$. Therefore, $P \obu_{3,1} = 0$ and $(i\wz I-L)\obu_{3,1} = \Phi \obu_{3,1}$, which implies that the field $\obu_{3,1}$ computed with $(i\wz I-L)$ instead of $\Phi$ is directly the particular solution of (\ref{eq:o3ls_af}). Note that $\obu_{3,1}$ appears as a true correction to $\hbu_o$ in the sense of the scalar product. This property has the striking and important consequence that the operator $\Phi$ never needs to be constructed explicitly, whatever the order of the amplitude equation. 
The homogeneous part of the solution of (\ref{eq:o3ls_af}) is arbitrarily proportional to $\hbu_o$. It can be ignored without loss of generality~\cite{Fujimura91}. Eventually, the term $P\obu_{j,1}e^{i\wz t} +c.c$ collected at order $O(\sqrt{\ez}^{j+2})$ disappears if $j\geq 2$. This is due to the nullity of $\obu_{j,1}$ for even $j$, and to the nullity of $P\obu_{j,1}$ for odd $j$. 
Overall, the particular solution at order $\sqrt{\ez}^3$ writes:
\begin{equation} 
\bm{u}_3(t,T) = \left( \phi \hbu_{3,1}^{(a)} + A \hbu_{3,1}^{(b)} + A\left | A \right |^2 \hbu_{3,1}^{(c)}  \right) e^{i\wz  t} +  A^3 e^{3 i\wz t} \hbu_{3,3} + c.c,
\end{equation}
where 
\begin{equation}
\begin{split}
(i\wz I-L) \hbu_{3,1}^{(a)} &= \hbf_h - \eta \gamma \hbu_o ,
\\
(i\wz I-L) \hbu_{3,1}^{(b)} &= - \hbf_o + \eta \hbu_o,
\\
(i\wz I-L) \hbu_{3,1}^{(c)} &= - 2C(\hbu_o,\bm{u}_{2,0}) - 2C(\hbu^*_o,\hbu_{2,2}) + \zeta \hbu_o 
\\
(3 i \wz I - L) \hbu_{3,3} &= - 2C(\hbu_o,\hbu_{2,2})
\end{split}
\end{equation}

The equation at order $\ez^2$ is assembled as: 
\begin{equation}
(\Phi \obu_{4,1}e^{i \wz t} + c.c) + \bm{s}_4  = - 2C(\bu_1,\bu_3) - C(\bu_2,\bu_2) - \partial_T \bm{u}_2 + (P \obu_{2,1}e^{i \wz t} + c.c)
\end{equation}
As mentioned, $P \obu_{2,1} = \bm{0}$ since $\obu_{2,1} = \bm{0}$, and the forcing terms are $-2C(\bu_1,\bu_3)$, $-C(\bu_2,\bu_2)$ and $-\pa_T \bu_2$. We first develop $C(\bu_1,\bu_3)$ as:
\begin{equation*}
\begin{split}
& C(\bu_1,\bu_3) = \phi A C(\hbu_o,\hbu_{3,1}^{(a)*})  + |A|^2  C(\hbu_o,\hbu_{3,1}^{(b)*}) + |A|^4  C(\hbu_o,\hbu_{3,1}^{(c)*})  \\
& + \left[ \phi A C(\hbu_o,\hbu_{3,1}^{(a)}) + A^2 C(\hbu_o,\hbu_{3,1}^{(b)}) + A^2 |A|^2 \left[ C(\hbu_o,\hbu_{3,1}^{(c)}) + C(\hbu_o^*,\hbu_{33}) \right]\right]e^{2i\wz t} \\
& + A^4 e^{4i\wz t} C(\hbu_o,\hbu_{33}) + c.c,
\end{split}
\end{equation*}
then $C(\bu_2,\bu_2)$ as: 
\begin{equation*}
\begin{split}
& C(\bu_2,\bu_2) =  |A|^4[C(\hbu_{2,2},\hbu_{2,2}^*) +c.c] + |A|^4 C(\bu_{2,0},\bu_{2,0})  \\
& + \left[ 2A^2 |A|^2 e^{2i\wz t}C(\bu_{2,0},\hbu_{2,2}) + c.c \right] + \left[ A^4 e^{4i\wz t} C(\hbu_{2,2},\hbu_{2,2}) + c.c\right] .
\end{split}
\end{equation*}
In addition, 
\begin{align*}
\pa_T |A|^2 &= A^*\pa_T A + A \pa_T A^* = A^*(\phi \eta - \eta A - \zeta A |A|^2) + A(\phi \eta^* - \eta^* A^* - \zeta^* A^* |A|^2) \\
& =  \phi \eta A^* + \phi \eta^* A  - (\eta + \eta^*) |A|^2 - (\zeta + \zeta^*) |A|^4,   
\end{align*}
and:
\begin{align*}
\pa_T A^2 & = 2 A \pa_T A = 2\phi \eta A  - 2 \eta A^2 - 2 \zeta A^2 |A|^2,
\end{align*}
such that: 
\begin{equation*}
\begin{split}
\pa_T \bu_2 & = \pa_T(|A|^2 \bu_{2,0}  +  A^2 e^{2i \wz t} \hbu_{2,2} + A^{*2} e^{-2i \wz t} \hbu_{2,2}^*) \\
& =(\phi \eta^* A \bu_{2,0}  + c.c ) - (\zeta + \zeta^*) |A|^4 \bu_{2,0} - (\eta + \eta^*)|A|^2\bu_{2,0} \\
& + [(2 \phi \eta  A \hbu_{2,2}  - 2 \eta A^2 \hbu_{2,2} - 2\zeta A^2 |A|^2\hbu_{2,2} )e^{2i\wz t} + c.c ].
\end{split}
\end{equation*}
Eventually, collecting all terms leads to the following particular solution for $\bu_4$: 
\begin{align*}
\bu_4 & = [\phi A \hbu_{4,0}^{(a)} + c.c] + |A|^2 \bu_{4,0}^{(b)} + |A|^4 \bu_{4,0}^{(c)} + ... \\
& [(\phi A\hbu_{4,2}^{(a)} + A^2 \hbu_{4,2}^{(b)} + A^2|A|^2 \hbu_{4,2}^{(c)} )e^{2i\wz t} + c.c] + [A^4e^{4i\wz t}\hbu_{4,4} + c.c],
\end{align*}
with:
\begin{align*}
-L \hbu_{4,0}^{(a)} &= -  \eta^* \bu_{2,0}  - 2C(\hbu_o,\hbu_{3,1}^{(a)*}),  \\
-L \bu_{4,0}^{(b)} &= \bu_{2,0}(\eta + \eta^*) - [ 2C(\hbu_o,\hbu_{3,1}^{(b)*}) + c.c ], \\
-L \bu_{4,0}^{(c)} &=\bu_{2,0}(\zeta + \zeta^*) -  C(\bu_{2,0},\bu_{2,0}) - [ C(\hbu_{2,2},\hbu_{2,2}^*) + c.c ] - [ 2C(\hbu_o,\hbu_{3,1}^{(c)*}) + c.c ],   \\ 
(2i\wz I - L) \hbu_{4,2}^{(a)} &= - 2\eta \hbu_{2,2} - 2C(\hbu_o,\hbu_{3,1}^{(a)}),  \\
(2i\wz I - L) \hbu_{4,2}^{(b)} &= 2\eta \hbu_{2,2} - 2C(\hbu_o,\hbu_{3,1}^{(b)}), \\ 
(2i\wz I - L) \hbu_{4,2}^{(c)} &= 2\zeta \hbu_{2,2} -  2C(\bu_{2,0},\hbu_{2,2}) - 2C(\hbu_o,\hbu_{3,1}^{(c)}) - 2C(\hbu_o^*,\hbu_{3,3}),  \\
(4i\wz I - L) \hbu_{4,4} &= -  C(\hbu_{2,2},\hbu_{2,2}) - 2C(\hbu_o,\hbu_{3,3}).
\end{align*}
The norm of the particular solutions at successive orders $\ez$, $\sqrt{\ez}^3$ and $\ez^{2}$ are outlined in table~\ref{tab:nord} for the plane Poiseuille flow at $(Re,k_x,k_z)=(3000,1.2,0)$ considered in \S\ref{sec:harmonicPoiseuille} and forced at $\wz=0.3810$. 
\begin{table}
  \begin{center}
  \def~{\hphantom{0}}
\begin{tabular}{cc}
 $||\bu_{2,0}||$ & $||\hbu_{2,2}||$ \\ \hline
 $11.5$ & $2.07$ \\
\end{tabular}
\\
\begin{tabular}{cccc}
 $||\hbu_{3,1}^{(a)}||$ & $||\hbu_{3,1}^{(b)}||$ & $||\hbu_{3,1}^{(c)}||$ & $||\hbu_{3,3}||$ \\ \hline
 $10.6$ & $10.6$ & $21.7$ & $11.4$ \\
\end{tabular}
\\
\begin{tabular}{cccccc}
 $||\hbu_{4,0}^{(a)}||$ & $||\bu_{4,0}^{(b)}||$ & $||\bu_{4,0}^{(c)}||$ & $||\hbu_{4,2}^{(a)}||$ & $||\hbu_{4,2}^{(b)}||$ & $||\hbu_{4,2}^{(c)}||$ \\ \hline
 $10.5\cdot 10^{4}$ & $4.2\cdot 10^{4}$ & $5.3\cdot 10^{4}$ & $1.6\cdot 10^{2}$ & $4.0\cdot 10^{2}$ & $1.6\cdot 10^{2}$ \\
\end{tabular}
\caption{Norms of the particular solutions at $\ord(\ez)$, $\ord(\sqrt{\ez}^3)$ and $\ord(\ez^{2})$ for the plane Poiseuille flow at $(Re,k_x,k_z)=(3000,1.2,0)$ considered in \S\ref{sec:harmonicPoiseuille}, and forced at $\wz=0.3810$ \label{tab:nord}}
\end{center}
\end{table}
Despite a large harmonic gain for $\w=0$ as visible in figure~\ref{fig:linpois}, the stationary field $\bu_{2,0}$ remains of reasonable amplitude as the associated Reynolds stress forcing $C(\hbu_o,\hbu_o^*)$ projects poorly on the most amplified singular mode for $\w=0$. However, the same does not hold for the stationary fields $\hbu_{4,0}^{(a,b,c)}$ at order $\ez^2$, all of significantly large amplitudes. This implies that the asymptotic hierarchy is only maintained until order $\sqrt{\ez}^{3}$. Indeed, if it holds that 
\begin{equation*}
\sqrt{\ez} \gg \ez(||\bu_{2,0}||,||\hbu_{2,2}||) \gg   \sqrt{\ez}^3 (||\hbu_{3,1}^{(a)}||,||\hbu_{3,1}^{(b)}||,||\hbu_{3,1}^{(c)}||,||\hbu_{3,3}||)
\end{equation*}
such that until order $\sqrt{\ez}^3$ each order appears as a true \textit{correction} of the previous one, this does not maintain for order $\ez^2$. As $\ez^2 ||\hbu_{4,0}^{(a)}||$ is of order unity, it cannot be considered as a correction of the order $\sqrt{\ez}^3$ but appears directly at the base flow level, which is asymptotically ill-posed.

\section{Modal amplitude equation for harmonic forcing. \label{appendix:Pois2}}

The dominant eigenmode $\hbq_1$ satisfies $L\hbq_1 = \sigma_1 \hbq_1$. Let its small damping rate $\sigma_{1,r}$ (i.e, the real part of $\sigma_1$), be scaled in terms of $\ez$ as $\sigma_{1,r} = \theta  \ez$, where $\theta = \ord(1)$ (and $\theta \leq 0$). The forcing frequency $\wz$ is detuned around the natural one, i.e, $\wz = \w_1 + \beta \ez$ where $\w_1$ is the imaginary part of $\sigma_1$ and $\beta = \ord(1)$. The shift-operator procedure introduced in \cite{Meliga12} is adopted thereafter, in order to apply the classical weakly nonlinear formalism. Namely, we perturb $L$ as $L = \bar{L} + \ez S$ where $S$ satisfies $S \hbq_1 = \theta \hbq_1$ and is such that all the others eigenvectors of $L$ constitute its kernel (i.e, $S\hbq_i=\bm{0}$ for $i=2,3,...$ ). In this way, the perturbed operator $\bar{L}$ possesses the same eigenvector as $L$, only the eigenvalue $\sigma_1$ associated with $\hbq_1$ is shifted of $-\sigma_{1,r}$ such as to be truly neutral: $\bar{L}\hbq_1 = (L-\ez S)\hbq_1 = L\hbq_1-\sigma_{1,r} \hbq_1 = i \w_1 \hbq_1$. The asymptotic multiple scale expansion of the forced Navier-Stokes equations expresses:
\begin{equation}
\begin{split}
& \sqrt{\ez} \Big[ (\partial_t-\bar{L})\bu_1 \Big] + \ez \Big[ (\partial_t-\bar{L})\bu_2 + C(\bu_1,\bu_1) \Big] + ...\\
& + \sqrt{\ez}^3 \Big[ (\partial_t - \bar{L})\bu_3 + 2C(\bu_1,\bu_2) + \partial_T \bm{u}_1 - S\bu_1 \Big] + O(\ez^2) = \phi \sqrt{\ez}^3 e^{i \wz t} \hbf_o  + c.c. 
\end{split}
\end{equation}
The equation at order $\sqrt{\ez}$ reads:
\begin{align*}
(\partial_t - \bar{L})\bu_1 = 0, 
\end{align*}
which leads to the solution $\bu_1 = A(T)\hbq_1e^{i \w_1 t} + c.c $. At order $\ez$, we obtain for $\bu_2$ the equation:
\begin{align*}
(\partial_t - \bar{L})\bu_2 &= -C(\bu_1,\bu_1) \\
& = -2|A|^2C(\hbq_1,\hbq_1^*) - \left[ A^2C(\hbq_1,\hbq_1)e^{i 2\w_1 t} + c.c \right],
\end{align*}
whose solution is $\bm{u}_2 = \left | A \right |^2 \bq_{2,0} + [ A^2 e ^{2i\wz t}\hbq_{2,2}  + c.c] $, where
\begin{align*}
 - \bar{L}  \bm{q}_{2,0} &=  - 2C(\hbq_1,\hbq^*_1),
 \\
(2 i   \wz I  - \bar{L})  \hbq_{2,2} &=  - C(\hbq_1,\hbq_1).
\end{align*}
At order $\sqrt{\ez}^3$ is assembled:
\begin{align*}
(\partial_t - \bar{L})\bu_3 &= -2C(\bu_1,\bu_2) + S \bu_1 - \pa_T \bu_1 + \phi( e^{i\beta T + i \w_1 t} \hbf_o + c.c) \\
& = \left[ -2A|A|^2 \left[C(\hbq_{2,2},\hbq_1^*) + C(\bq_{2,0}, \hbq_1)\right] + \theta A \hbq_1  - \hbq_1\frac{\mathrm{d} A}{\mathrm{d} T} +  \phi e^{i \beta T} \hbf_o \right]e^{i \w_1 t}  \\
& + c.c + \text{\textit{non-resonant terms}}
\end{align*}
where we used that $S\bu_1 = \theta A \hbq_1e^{i \w_1 t} + c.c$. Canceling the projection of the resonant part of the forcing term (inside the brackets) on the adjoint $\hba_1$, results in an equation for $A$: 
\begin{align*}
\frac{\mathrm{d} A}{\mathrm{d} T} = \theta A - A|A|^2\frac{\langle 2C(\hbq_{2,2},\hbq_1^*) + 2C(\bq_{2,0}, \hbq_1), \hba_1 \rangle }{\langle \hbq_1 , \hba_1\rangle} + \phi e^{i\beta T} \frac{\langle \hbf_o , \hba_1 \rangle}{\langle \hbq_1 , \hba_1 \rangle}. 
\end{align*}
Note that for $\wz = \w_1$ (i.e, the detuning parameter $\beta=0$) the amplitude in the linear regime, $A_l$, reads: 
\begin{align*}
0 = \theta A + \phi \frac{\langle \hbf_o , \hba_1 \rangle}{\langle \hbq_1 , \hba_1 \rangle} \Leftrightarrow A_l = -\phi \frac{\langle \hbf_o , \hba_1 \rangle}{\langle \hbq_1 , \hba_1 \rangle} \theta^{-1},
\end{align*}
which corresponds to the following linear harmonic gain:
\begin{align*}
G = \frac{\sqrt{\ez} |A_l| }{\phi \sqrt{\ez}^3} = \frac{1}{\ez}  \left |\frac{\langle \hbf_o , \hba_1 \rangle}{\langle \hbq_1 , \hba_1 \rangle} \right |\frac{\ez}{|\sigma_{1,r}|} =  \left| \frac{\langle \hbf_o , \hba_1 \rangle}{\langle \hbq_1 , \hba_1 \rangle} \right |\frac{1}{|\sigma_{1,r}|},
\end{align*}
which is different from the norm of the resolvent operator, i.e, $1/\ez$. Thus, even the matching with the linear regime is not guaranteed with this classical, modal approach. 
\section{Higher-order corrections of the WNNt equation.\label{appendix:TGFuji}}

Recall the equation (\ref{eq:tgo2}) obtained at order $\ez^2$:
\begin{equation*}
\Phi \bm{u}_2  = \alpha \bu_o 
+ A^2 e^{-L t}   \forcutwo
- \frac{\mathrm{d} A }{\mathrm{d} T} \ez \bu_o t -  A \bu_o.
\end{equation*}
After satisfaction of the Fredholm alternative, which leads to an equation for the amplitude $A$, the relation (\ref{eq:tgo2}) can be re-expressed as
\begin{equation}
\begin{split}
\Phi \bu_2 = A^2 \left( e^{-L t} \forcutwo - \ez \mu_2 \bu_o \right) \quad \text{for} \quad t>0, 
\end{split}
\label{eq:w2af}
\end{equation}
where the orthogonality of the right-hand side (RHS) with $\ba(t)$ is ensured by construction of $\mu_2(t) = \left \langle \forcutwo(t) , \bl(t) \right \rangle/\left \langle \bl(t) , \bl(t) \right \rangle$ in (\ref{eq:mu2}).
 The general solution to (\ref{eq:w2af}) reads $\bu_2 = \bu_2^{(\perp \bl)} + A_2 \bl(t)$. The particular solution of the system, $\bu_2^{(\perp \bl)}$, is obtained by solving (\ref{eq:w2af}) after replacing $\Phi(t)$ by $e^{-L t}$. 
 Indeed, such $\bu_2^{(\perp \bl)}$ must be orthogonal to $\bl(t)$, since $0 = \left \langle \mbox{RHS}, \ba(t) \right \rangle$  = $\left \langle  e^{Lt}(\mbox{RHS}), \bm{l}(t) \right \rangle$ = $\left \langle \bu_2^{(\perp \bl)}, \bm{l}(t) \right \rangle$. 
On the other hand, the term $A_2 \bl(t)$ constitutes the homogeneous part of the solution, where $A_2$ is a scalar amplitude. It can be kept in further calculations, provided it is included in the final amplitude for $\bl(t)$, which would then become $\ez A + \ez^2A_2 + O(\ez^3)$. Instead, and without loss of generality (\cite{Fujimura91}), we propose to set $A_2=0$ such that 
\begin{equation}
\begin{split}
\bu_2 = \bu_2^{(\perp \bl)} = A^2 \left(\forcutwo - \mu_2 \bl \right) \quad \text{for} \quad t>0.
\end{split}
\end{equation}
In particular, this implies that the term $\partial_t(P \bu_2)$ that appears at order $O(\ez^3)$ actually vanishes since $P\bu_2=P\bu_2^{(\perp \bl)}=\bm{0}$. If this is performed at each order $j \geq 3$, all the terms $\partial_t(P \bu_j)$  vanish. In this way, the ''retroaction'' forcing due to the operator perturbation only appears at $O(\ez^2)$.

Deriving a higher-order amplitude equation for transient growth requires introducing a very long time scale $\tau = \ez^2 t$, such that $A = A(T,\tau)$. The total derivative in $T$ should then replaced as partial derivative, and the amplitude equation derived at order $\ez^2$ writes $\pa_T A = A^2\dot{\mu}_2$ subject to $\lim_{t\rightarrow 0} ( \alpha - A ) =0 $. One gathers at order $\ez^3$ for $t>0$:
\begin{align}
 \partial_t (\Phi \bu_3)  =&  - 2A^3e^{-L t}C(\bl,\bu_2) - e^{-L t} \partial_T \bu_2 - e^{-L t} \partial_\tau \bu_1 - \partial_t(P\bu_2) \nonumber\\
=& - 2A^3He^{-L t}\left[ C(\bl,\forcutwo) -\mu_2 C(\bl,\bl)\right]  - 2 A (\partial_T A )\left(e^{-L t}\forcutwo - \ez \mu_2 \bu_o \right) - \ez \bu_o \partial_{\tau} A \nonumber \\
=& - 2A^3 e^{-L t}\left[ C(\bl,\forcutwo) - \mu_2 C(\bl,\bl) + \dot{\mu}_2\left(\forcutwo - \mu_2  \bl \right)\right] - \ez \bu_o \partial_{\tau} A, \label{eq:o3td}
\end{align}
since (i) $\partial_T \bu_2 = 2 A (\partial_T A )\left(\forcutwo - \mu_2 \bl \right)$, (ii) $P\bu_2 = \bm{0}$, and (iii) $e^{-L t}\partial_\tau \bu_1 = H e^{-L t} \bl \partial_{\tau} A = H \ez \bu_o \partial_{\tau}A$. Equation (\ref{eq:o3td}) is subject to $\bu_3(0) = \bm{0}$. Its particular solution yields
\begin{equation}
\bu_3(t,T,\tau)= A(T,\tau)^3\bu_3^{(a)}(t) + \frac{\partial A(T,\tau)}{\partial \tau}\bu_3^{(b)}(t),
\end{equation}
where
\begin{align*}
&\mathrm{d}_t\left[\Phi \bu_3^{(a)}\right] = - 2e^{-L t}\left[C(\bl,\forcutwo) - \mu_2 C(\bl,\bl) + \dot{\mu}_2(\forcutwo- \mu_2 \bl) \right], \quad \text{and} \quad \\
&\mathrm{d}_t\left[\Phi \bu_3^{(b)}\right] = - \ez \bu_o, 
\end{align*}
subject to the initial conditions $\bu_3^{(a)}(0) = \bu_3^{(b)}(0) = \bm{0}$. After time integration, we obtain
\begin{align}
\Phi \bu_3  &= A^3 e^{-L t} \forcuthree - \ez \bu_o t \partial_{\tau} A, \quad t>0,
\label{eq:tgo3}
\end{align}
where $\forcuthree$ is solution of
\begin{equation*}
\frac{\mathrm{d} \forcuthree}{\mathrm{d} t}  = L\forcuthree - 2\left[C(\bl,\forcutwo) - \mu_2 C(\bl,\bl) + \dot{\mu}_2(\forcutwo- \mu_2\bl) \right], \quad  \forcuthree(0) = \bm{0}. 
\end{equation*}
Canceling the projection of the RHS of  (\ref{eq:tgo3}) on $\ba(t)$, dividing the ensuing relation by $\left \langle \ba(t),\bu_o \right \rangle$, and taking the partial derivative with respect to $t$ leads to  
\begin{equation}
\ez A^3\frac{\mathrm{d} \mu_3}{\mathrm{d} t}  =  \ez \frac{\partial A}{\partial \tau}, \quad \quad t>0,
\label{eq:atg3}
\end{equation}
where
\begin{equation*}
\mu_3(t) = \ez^{-1}\frac{\left \langle  e^{-L t} \forcuthree(t) , \ba(t)  \right \rangle}{\left \langle  \bu_o , \ba(t)  \right \rangle} = \frac{\left \langle  \forcuthree(t) , \bl(t)  \right \rangle}{\left \langle  \bl(t) , \bl(t)  \right \rangle}.
\end{equation*}
%
To be meaningful, equation (\ref{eq:atg3}) and its initial condition must be re-written solely in terms of $t$, which is done by evaluating $T=\ez t$ and $\tau = \ez^2t$. The \textit{total} derivative of $A$, denoted $d_t A$, is now needed, as it takes into account the \textit{implicit} dependence of $A$; it reads $\mathrm{d}_t A = \partial_t A + \ez \partial_T A + \ez^2 \partial_{\tau} A = \ez \partial_T A + \ez^2 \partial_{\tau} A$, such that
\begin{equation*}
\ez A^2\frac{\mathrm{d} \mu_2(t) }{\mathrm{d} t} + \ez^2 A^3\frac{\mathrm{d} \mu_3(t) }{\mathrm{d} t}  =  \frac{\mathrm{d} A}{\mathrm{d} t}, \quad  t>0,
\end{equation*}
subject to $\lim_{t\rightarrow 0} ( \alpha - A(\ez t,\ez^2t) ) =0$ so $A(t \rightarrow 0) = \alpha$ and the amplitude $A$ is extended by continuity in $t=0$ so as to  impose $A(0) = \alpha$. 

\section{Transient gain sensitivity and comparison with the WNNt model.\label{appendix:S11}}

We consider a linear system $\partial_t \bu = L\bu$ subject to the initial condition $\bu(0)$ with $||\bu(0)||=1$. The linear transient gain at $t=t_o$ writes $G_o = ||\bu(t_o)||$. A variational method is used to 
derive the variation of the optimal transient gain induced by a small perturbation $\de L$ of operator. Let us introduce the Lagrangian
\begin{equation*}
\mL = G_o^2 - \int_{0}^{t_o} \left \langle \partial_t \bu - L\bu , \bu^\da \right \rangle \mathrm{d} t - \beta \left( 1-||\bu(0)||^2 \right),
\end{equation*}
where the  Lagrange multipliers $\bu^\da$ and $\beta$ enforce the constraints on the state equation and on the norm of the initial condition, respectively.
Imposing $\left \langle \partial_{\bu} \mL , \de \bu \right \rangle = 0$ for all $\de \bu$ 
leads to the adjoint equation 
$\partial_t \bu^\da = - L^\da \bu^\da$, to be integrated backward in time from the terminal condition  $\bu^\da(t_o) = 2\bu(t_o)$. Eventually, the gain variation induced by $\de L$ is
\begin{equation}
\de (G_o^2) = \left \langle \frac{\partial \mL }{\partial L} , \de L \right \rangle = \int_{0}^{t_o} \left \langle (\de L) \bu , \bu^\da \right \rangle \mathrm{d}t.
\label{eq:tgsensi}
\end{equation}
On the other hand, we derived in the main text for the WNNt model:  
\begin{align}
a(t) &= \frac{U_0}{\ez}\left[1-\left( \frac{U_0}{\ez} \right)^2 2\mu_3(t) \right]^{-1/2}.
\label{eq:atg}
\end{align}
The weakly-nonlinear transient gain squared can be expressed as $G(t_o)^2=(a(t_o)/U_0)^2$, while the linear gain squared is $G_o^2 = (1/\ez)^2$, such that
\begin{align*}
\frac{1}{G(t_o)^2} -\frac{1}{G_o^2} &= -U_0^2 2 \mu_3(t_o). 
\end{align*}
We are interested in small variations around $G_o^2$, thus we write $G(t_o)^2 = G_o^2 + \de G_o^2$ with $|\de G_o^2/G_o^2| \ll 1$. In this manner, $1/G_o^2-1/G(t_o)^2 = \de (G_o^2 )/G_o^4 + h.o.t$, eventually leading to 
\begin{align*}
\de (G_o^2 ) &= 2 \mu_3(t_o) \frac{U_0^2}{\ez^4}. 
\end{align*}
In addition,
\begin{align*}
\mu_3(t_o) = \ez^{-1} \frac{\left \langle e^{-Lt_o} \forcuthree(t_o), \ba(t_o) \right \rangle}{\left \langle \bu_o , \ba(t_o)\right \rangle},
\end{align*}
with $\forcuthree(t) = - e^{Lt}\int_{0}^{t}2e^{-Ls}C[\forcutwo(s),\bl(s)] \md s $. 
Therefore
\begin{align*}
\mu_3(t_o) & = - \ez^{-1} \frac{\left \langle \int_{0}^{t_o}2e^{-Ls}C[\forcutwo(s),\bl(s)] \md s, \ba(t_o) \right \rangle}{\left \langle \bu_o , \ba(t_o)\right \rangle}  \\
 & = - \ez^{-1} \frac{\int_{0}^{t_o} \left \langle 2e^{-Ls}C[\forcutwo(s),\bl(s)] , \ba(t_o)  \right \rangle \md s }{\left \langle \bu_o , \ba(t_o)\right \rangle}.
\end{align*}
By definition, $\ba(t_o) = (e^{L t_o})^\da \bl(t_o)$ and $\left \langle \bu_o , \ba(t_o)\right \rangle = \left \langle e^{Lt_o}\bu_o , \bl(t_o)\right \rangle = 1/\ez $.
In addition, $\left( e^{-Ls} \right)^\da(e^{L t_o})^\da = \left(e^{L t_o}e^{-Ls} \right)^\da = \left(e^{-L(s- t_o)}\right)^\da$, such that
\begin{align*}
\mu_3(t_o) & = - \int_{0}^{t_o} \left \langle 2C[\forcutwo(s),\bl(s)] , \left(e^{-L(s- t_o)}\right)^\da\bl(t_o)  \right \rangle \md s. 
\end{align*}
In terms of our previous notations, we have the direct correspondence  $\bu(t) = \bl(t)/\ez$ and $\bu^{\dagger}(s) =  \left(e^{-L(s- t_o)}\right)^\da 2 \bu(t_o)$, so we can write 
\begin{align*}
\delta (G_o^2) & = \left(\frac{2U_0^2}{\ez^4}\right)\left( - \frac{\ez^2}{2} \int_{0}^{t_o} \left \langle 2C[\forcutwo(s),\bu(s)] , \left(e^{-L(s- t_o)}\right)^\da 2\bu(t_o)  \right \rangle \md s\right) \\
& = - \int_{0}^{t_o} \left \langle 2C\left[(U_0/\ez)^2\forcutwo(s),\bu(s)\right] , \bu^{\dagger}(s) \right \rangle \md s.
\end{align*}
The sensitivity relation (\ref{eq:tgsensi}) is immediately recognised, where $\de L$ is here induced by the addition of $(U_0/\ez)^2\forcutwo$ to the base flow. Indeed, $U_0/\ez = a_{lin}$ is the linear solution of (\ref{eq:atg}) corresponding the limit $U_0 \rightarrow 0$, such that the flow field is described in this limit by $\bm{U}(t) = \bm{U}_e + a_{lin}\bl(t) + a_{lin}^2\forcutwo(t) + O(\ez^3)$.

\bibliographystyle{jfm}
\bibliography{jfm-instructions}

\end{document}